\documentclass[american]{revtex4-1}
\usepackage[T1]{fontenc}
\usepackage[latin9]{inputenc}
\usepackage{geometry}
\usepackage{babel}
\usepackage{amsmath}
\usepackage{amsthm}
\usepackage{amssymb}
\usepackage{graphicx}
\usepackage[unicode=true]
 {hyperref}

\makeatletter

\providecommand{\tabularnewline}{\\}

\makeatother

\begin{document}

\title{Markov State Models from short non-Equilibrium Simulations - Analysis
and Correction of Estimation Bias}

\author{Feliks Nüske}

\author{Hao Wu}

\author{Jan-Hendrik Prinz}

\author{Christoph Wehmeyer}

\author{Cecilia Clementi}

\author{Frank Noé}

\affiliation{Freie Universität Berlin, Department of Mathematics and Computer
Science, Arnimallee 6, 14195 Berlin, Germany}

\affiliation{Rice University, Center for Theoretical Biological Physics, and Department
of Chemistry, Houston, Texas, 77005, United States}
\email{feliks.nueske@fu-berlin.de}

\email{frank.noe@fu-berlin.de}

\begin{abstract}
Many state of the art methods for the thermodynamic and kinetic characterization
of large and complex biomolecular systems by simulation rely on ensemble
approaches, where data from large numbers of relatively short trajectories
are integrated. In this context, Markov state models (MSMs) are extremely
popular because they can be used to compute stationary quantities
and long-time kinetics from ensembles of short simulations, provided
that these short simulations are in ``local equilibrium'' within
the MSM states. However, in the last over 15 years since the inception
of MSMs, it has been controversially discussed and not yet been answered
how deviations from local equilibrium can be detected, whether these
deviations induce a practical bias in MSM estimation, and how to correct
for them. In this paper, we address these issues: We systematically
analyze the estimation of Markov state models (MSMs) from short non-equilibrium
simulations, and we provide an expression for the error between unbiased
transition probabilities and the expected estimate from many short
simulations. We show that the unbiased MSM estimate can be obtained
even from relatively short non-equilibrium simulations in the limit
of long lag times and good discretization. Further, we exploit observable
operator model (OOM) theory to derive an unbiased estimator for the
MSM transition matrix that corrects for the effect of starting out
of equilibrium, even when short lag times are used. Finally, we show
how the OOM framework can be used to estimate the exact eigenvalues
or relaxation timescales of the system without estimating an MSM transition
matrix, which allows us to practically assess the discretization quality
of the MSM. Applications to model systems and molecular dynamics simulation
data of alanine dipeptide are included for illustration. The improved
MSM estimator is implemented in PyEMMA as of version 2.3.
\end{abstract}
\maketitle

\section{Introduction}

Ensemble approaches, where many fairly short simulations are produced
in parallel or on distributed computer architectures, are widely used
in order to characterize the thermodynamics and kinetics of large
biological macromolecules., Markov state models (MSMs) \cite{PrinzEtAl_JCP10_MSM1,Schutte:2013aa,MSMBook:2014aa}
have become standard tools for the analysis of such data sets generated
by molecular dynamics (MD) simulations \cite{SchuetteFischerHuisingaDeuflhard_JCompPhys151_146,Swope:2004aa,NoeHorenkeSchutteSmith_JCP07_Metastability,ChoderaEtAl_JCP07,Noe:2008aa,Buchete:2008aa,Bowman_JCP09_Villin,NoeSchuetteReichWeikl_PNAS09_TPT,Voelz:2010aa,Zhuang_JPCB11_MSM-IR}.
An MSM provides a simplified model of the underlying Markov process,
which is continuous in both time and space, by a discrete time Markov
chain on finitely many states. These states are defined by partitioning
the continuous state space into finitely many disjoint sets. Time
is discretized by choosing a discrete time step, called the \textit{lag
time}, and the full process is replaced by a snapshot process that
only keeps track of the discrete state visited at the discrete time
steps, discarding any time information in between and any spatial
information within the discrete sets. The quality of this approximation
critically depends on the choice of both discretization and lag time
\cite{SarichNoeSchuette_MMS09_MSMerror}. One of the strengths of
Markov models is that the simulations used to construct them do not
necessarily need to sample from the global equilibrium distribution,
as only conditional transition probabilities between the states are
required \cite{SchuetteFischerHuisingaDeuflhard_JCompPhys151_146}.
In particular, at least in principle, these transition probabilities
can be obtained without bias from simulations started out of local
equilibrium in each state which only run for the length of a single
lag time step. However, it is much more practical to produce simulations
that are longer than one lag time and estimate MSMs by counting transitions
along these trajectories. Even if the simulations are started out
of local equilibrium, the distribution deviates from local equilibrium
over time until global equilibrium is restored. The estimation of
transition probabilities is therefore subjected to a bias \cite{PrinzEtAl_JCP10_MSM1}.
In order to keep the bias small, it must be assumed that local equilibrium
is approximately restored after every time step.

The effect of the initial distribution onto the MSM quality or even
the justification of using an MSM for data analysis has been controversially
discussed, and this issue has not been resolved yet. At least three
ideas have been discussed \cite{Chodera:aa}: (1) This effect exists
\cite{PrinzEtAl_JCP10_MSM1}, but may be small and can be ignored
in practice. (2) We can reduce the effect of non-equilibrium starting
points by discarding the first bit of simulation trajectories, enough
to reach local equilibrium \cite{NoeSchuetteReichWeikl_PNAS09_TPT}.
(3) We can avoid this problem by preparing local equilibrium distributions
in the starting states using biased simulations and then shooting
trajectories out of them \cite{Roeblitz_PhD,Weber:2009aa,Huang:2009aa,Bowman:2009aa}.

Here we qualify and quantify these ideas by systematically analyzing
the effect of non-equilibrium starting conditions onto MSM quality,
and we suggest effective correction mechanisms. Throughout the manuscript,
we use the term ``non-equilibrium'' to describe the problem that
simulations are started from a distribution which is not in global
equilibrium, and their simulation time is too short to reach that
global equilibrium. Note, however, that the dynamics itself is assumed
to possess a unique equilibrium distribution, and if long enough simulations
would be run, they would sample from the equilibrium distribution.
Briefly, our main results are:
\begin{enumerate}
\item We provide an expression for the error between unbiased transition
probabilities and the expected estimate from many simulations running
for multiple discrete time steps, see Section \ref{sec:Estimation_Error}.
We find that there is no fundamental advantage of starting simulations
in local equilibrium. Rather, the estimation error depends on the
discretization, the simulation length and the lag time. In the limit
of long lag times and fine discretization, MSMs are estimated without
bias even when non-equilibrium starting points are used. However,
for a given discretization the lag time required to practically achieve
a small estimation bias might be large.
\item We derive an unbiased MSM estimator that corrects the error due to
non-equilibrium starting conditions at short lag times, by exploiting
the framework of \textit{observable operator models} (OOMs) - see
Sec. \ref{sec:OOMs}. OOMs are powerful finite-dimensional models
that provide unbiased estimates of stationary and kinetic properties
of stochastic processes under fairly mild assumptions, see \cite{Jaeger:2000aa,Wu:2015aa,Wu:2016aa}.
Most importantly, OOMs can be estimated from non-equilibrium simulations
\cite{Wu:2016aa} and are not limited to a local equilibrium assumption. 
\item We utilize the fact that exact relaxation timescales that are not
contaminated by the MSM projection error (i.e. quality of the coordinates
and the clustering used) can be estimated using the OOM framework.
The difference between the unbiased estimate and the uncorrected or
corrected MSM estimate is very insightful as it provides an indicator
of the quality of the MSM discretization. If this difference is too
large, it is suggested to rather improve the coordinate selection
or discretization used for MSM construction and re-analyze. Note that
while OOMs offer the more general theory, they are not as easy to
interpret and their estimation from finite data is not as stable and
mature as MSM estimation.
\end{enumerate}
As a technical advance, we provide a meaningful strategy to select
the model rank of an OOM which is required in order to obtain practically
useful estimates, by using a statistical analysis of singular values
of the count matrix (Sec. \ref{subsec:Numerical_Rank}). 

Sec. \ref{sec:Examples}, demonstrates the usefulness of the OOM framework
for two model systems and MD simulation data of alanine dipeptide.
We show that accurate estimates of spectral and stationary properties
can be obtained from short non-equilibrium simulations, even for short
lag times or poor discretizations. We explain how the discretization
quality is revealed by the difference between spectral estimates of
MSM and OOM. We also show that the rank selection strategy helps to
choose a suitable model rank even for small lag times, when no apparent
timescale separation can be utilized.

\begin{figure}
\begin{centering}
\includegraphics{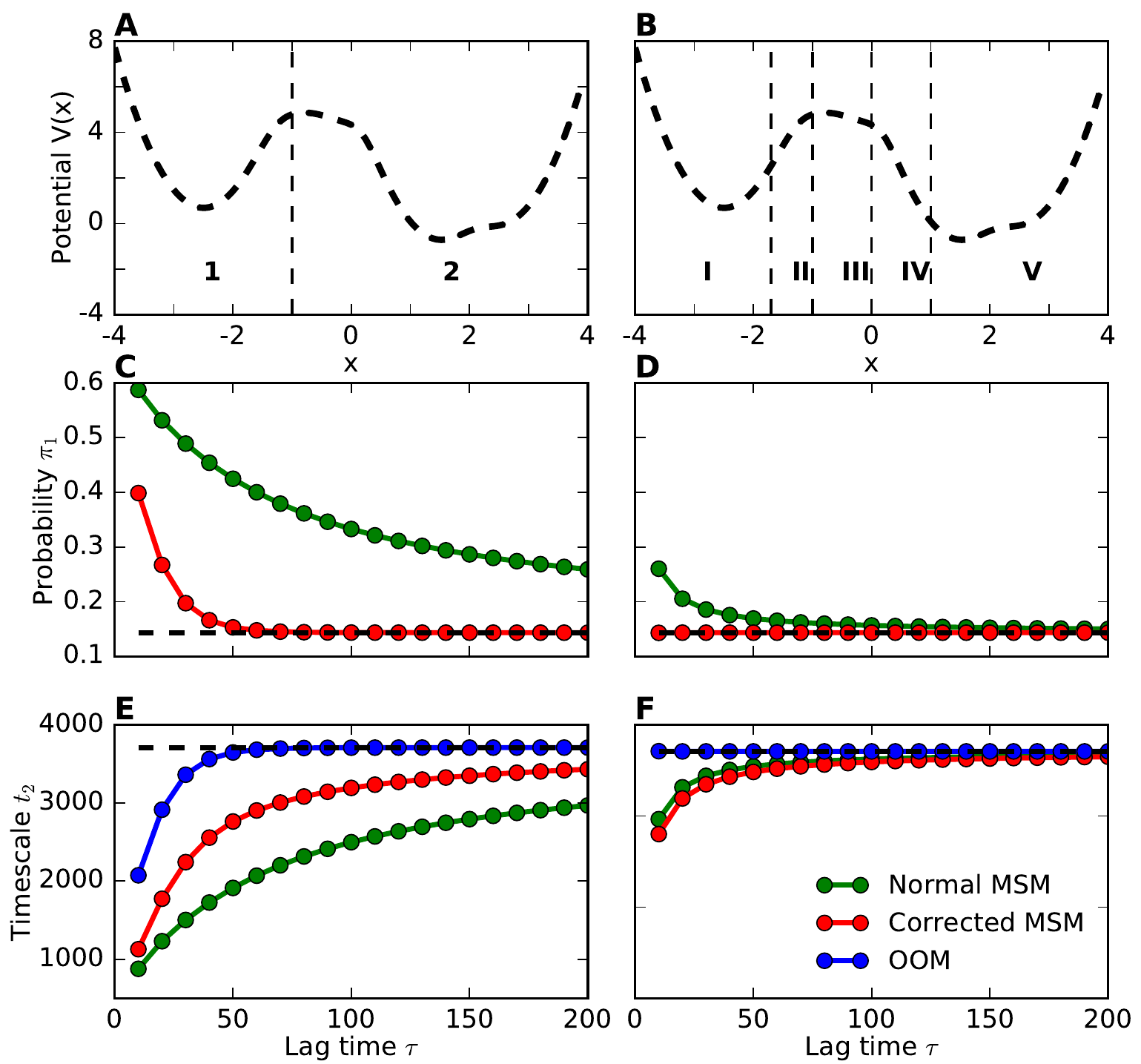}
\par\end{centering}
\caption{A: One-dimensional potential function and discretization into two
states. B: The same potential with a five state discretization. C,
D: Estimates for the equilibrium probability of state 1 from the direct
MSM (green) and the unbiased MSM (red), reference in black. E, F:
Estimates for the slowest relaxation timescale $t_{2}$ from a direct
MSM (green), c.f. Eq. (\ref{eq:Empirical_Mean_TransitionMatrix}),
the unbiased MSM (red), c.f. Eqs. (\ref{eq:OOM_Path_Prob_C_Eq}-\ref{eq:OOM_Path_Prob_pi_i}),
and the spectral OOM estimation (blue), Eqs. (\ref{eq:EV_Decomp_E_Omega_0}-\ref{eq:EV_Decomp_E_Omega_1}).
The black dashed line corresponds to the reference value. \label{fig:intro_figure}}
\end{figure}

As an illustration, consider the one-dimensional model system governed
by the potential shown in Fig. \ref{fig:intro_figure} A, see Sec.
\ref{subsec:One_d_model_system} for details. We study the estimation
of a Markov model using the two state discretization indicated in
panel A of Fig. \ref{fig:intro_figure}. For various lag times, we
investigate the expected transition matrix if 90 per cent of the simulations
are started from local equilibrium within state 1, while the other
10 per cent are started from local equilibrium within state 2. Note
that we do not use any simulation data here, we only compute expected
values over an ensemble of trajectories, with the trajectory length
set to $2000$ steps, which is shorter than the slowest relaxation
timescale.

For short lag times, the standard MSM provides a strongly biased estimate
of the equilibrium population of the two wells (Fig. \ref{fig:intro_figure}C,
green curve). For longer lag times, the MSM converges towards the
correct equilibrium population, but the bias only disappears when
the lag time approaches the longest relaxation timescale of the system,
so if the initial distribution is far from equilibrium this can entail
a significant error at practically feasible lag times. In contrast,
the corrected MSM estimate proposed in this paper achieves the correct
estimate of equilibrium populations even at short lag times (Fig.
\ref{fig:intro_figure}C, red curve). The standard MSM relaxation
timescales are underestimated at short lag times, consistent with
previous variational results \cite{Djurdjevac:2012aa,NoeNueske_MMS13_VariationalApproach,Nuske:2014aa},
but they can be improved by using the unbiased MSM estimator proposed
here (Fig. \ref{fig:intro_figure}E). The OOM can provide a model-free
estimate of the relaxation timescale that is unbiased at a relatively
short lag time (Fig. \ref{fig:intro_figure}E, blue line). The difference
between the OOM and the corrected MSM estimate (blue versus red lines
in Fig. \ref{fig:intro_figure}E) is an indicator of the MSM model
error due to the state space discretization. Please note that all
MSM results in this figure can be dramatically improved if a finer
clustering is used. For example, if the five state partitioning from
Fig. \ref{fig:intro_figure}B is used instead, the estimation of stationary
properties converges much faster (Fig. \ref{fig:intro_figure}D),
and there is hardly a difference between the timescales estimated
by a direct and an unbiased MSM (Fig. \ref{fig:intro_figure}F).

\section{Analysis of the State of the Art: MSM Estimation from Simulations
with arbitrary Starting Points}

\label{sec:Estimation_Error}

\subsection{Molecular Dynamics, Count Matrix and Transition Matrix}

In this work, we consider the setting described in detail in Ref.
\cite{PrinzEtAl_JCP10_MSM1}, that is, an ergodic and reversible Markov
process $X_{t}$ on continuous state space $\Omega$, which possesses
a unique stationary distribution $\pi$. We denote by $\tau>0$ the
lag time and by
\begin{eqnarray}
p(x,y;\tau) & = & \mathbb{P}(X_{\tau}\in dy|X_{0}=x)\label{eq:TransitionKernel}
\end{eqnarray}
the conditional transition density function, that is the probability
that the process, when located at configuration $x$ at time $t$,
will be found at configuration $y$ at time $t+\tau$. The corresponding
transfer operator is denoted by $\mathcal{T}(\tau)$ and is defined
by its action on a function of state space $u$:

\begin{eqnarray}
\mathcal{T}(\tau)u(y) & = & \int_{\Omega}p(x,y;\tau)\frac{\pi(x)}{\pi(y)}u(x)\,\mathrm{dx}.\label{eq:Definition_Transfer_Operator}
\end{eqnarray}

Its eigenvalues are called

\begin{eqnarray}
\lambda_{m}(\tau) & = & \exp(-\tau/t_{m}),\label{eq:Def_timescales}
\end{eqnarray}
where $t_{m}$ are the implied relaxation timescales. We denote the
transfer operator eigenfunctions by $\psi_{m},\,m=1,\ldots$ In particular,
we have that $\psi_{1}\equiv1$. If the transfer operator is of rank
$M$ at lag time $\tau$, the transition density can be written as
\begin{eqnarray}
p(x,y;\tau) & = & \sum_{m=1}^{M}\lambda_{m}(\tau)\,\psi_{m}(x)\,\pi(y)\,\psi_{m}(y).\label{eq:SpectralDecompKernel}
\end{eqnarray}
Note that exact equality in Eq. (\ref{eq:SpectralDecompKernel}) is
an assumption, but often it is satisfied approximately for a large
range of lag times $\tau$. Throughout the paper, we will consider
decompositions of state space into disjoint sets $S_{1},...,S_{N}$,
where $\Omega=\bigcup_{i}S_{i}$. The indicator function of set $S_{i}$
is called $\chi_{i}$. For a simulation of the continuous dynamics
which samples positions at discrete time steps, we will denote the
position at the $k$-th time step by $X_{k},\,k=1,\ldots,K$, s.t.
$K$ is the total number of time steps in the simulation. We use the
symbol $\mathbf{Y}$ as a shorthand notation for an entire simulation.
If multiple different simulations need to be distinguished, we will
denote them by $\mathbf{Y}_{q},\,q=1,\ldots,Q$, i.e. $Q$ is the
total number of available simulations.

Most of this work is based on correlations between the discrete sets.
For a trajectory as above, we define the empirical histograms and
correlations (also called state-to-state time-correlations) as follows:
\begin{eqnarray}
\mathbf{s}(i) & := & \frac{1}{K-2\tau}\sum_{k=1}^{K-2\tau}\chi_{i}(X_{k}),\label{eq:Definition_si}\\
\mathbf{S}^{\tau}(i,j) & := & \frac{1}{K-2\tau}\sum_{k=1}^{K-2\tau}\chi_{i}(X_{k})\chi_{j}(X_{k+\tau}),\label{eq:Definition_Stau}\\
\mathbf{S}_{r}^{2\tau}(i,j) & := & \frac{1}{K-2\tau}\sum_{k=1}^{K-2\tau}\chi_{i}(X_{k})\chi_{r}(X_{k+\tau})\chi_{j}(X_{k+2\tau}).\label{eq:Definition_S2tau}
\end{eqnarray}
Up to the normalization, the matrix $\mathbf{S}^{\tau}\in\mathbb{R}^{N\times N}$
is a \textit{count matrix} because it simply counts the number of
transitions from state $S_{i}$ to $S_{j}$ over a time window $\tau$
that have occurred in the simulation, while the vector $\mathbf{s}\in\mathbb{R}^{N}$
counts the total visits to state $S_{i}$ and corresponds to the $i$-th
row sum of $\mathbf{S}^{\tau}$. For each set $S_{r}$, the matrix
$\mathbf{S}_{r}^{2\tau}\in\mathbb{R}^{N\times N}$ is proportional
to a two-step count matrix counting subsequent transitions from state
$S_{i}$ to $S_{r}$ and on to state $S_{j}$. At first sight, it
may seem confusing that $\mathbf{S}^{\tau}$ and $\mathbf{s}$ only
count transitions and visits up to time $K-2\tau$, but further below,
we will use all three matrices in conjunction which requires estimating
all of them over the same part of the data. We will continue to refer
to these matrices as count matrix, count vector and two-step count
matrix in what follows. Also note that in the literature, the count
matrix and vector are often denoted by $\mathbf{C}^{\tau},\,\mathbf{c}$,
but we will use these symbols differently in what follows. Let us
note at this point that $\mathbf{s},\,\mathbf{S}^{\tau},\,\mathbf{S}_{r}^{2\tau}$
can be seen as random variables that map a (stochastic) trajectory
$\mathbf{Y}$ of discrete time steps to the values given in Eqs. (\ref{eq:Definition_si}-\ref{eq:Definition_S2tau}).
To emphasize this dependence, we will also write $\mathbf{s}(\mathcal{\mathbf{Y}}),\,\mathbf{S}^{\tau}(\mathbf{Y}),\,\mathbf{S}_{r}^{2\tau}(\mathbf{Y})$
if appropriate.

We are concerned with the estimation of a transition probability matrix
between the sets $S_{i}$ of a given discretization of state space.
If the process is in equilibrium, the conditional transition probabilities
can be expressed as
\begin{eqnarray}
\mathbf{T}_{Eq}^{\tau}(i,j) & = & \frac{\mathbb{P}\left(X_{t}\in S_{i},\,X_{t+\tau}\in S_{j}\right)}{\mathbb{P}(X_{t}\in S_{i})}\label{eq:stationary_pij_0}\\
 & = & \frac{\int_{S_{i}}\mathrm{dx}\int_{S_{j}}\mathrm{dy}\,\pi(x)\,p(x,y;\tau)}{\int_{S_{i}}\mathrm{dx}\,\pi(x)}\label{eq:stationary_pij_1}\\
 & = & \frac{\mathbf{C}_{Eq}^{\tau}(i,j)}{\pi_{i}}.\label{eq:stationary_pij_2}
\end{eqnarray}
Here, we have defined the \textit{equilibrium correlation} between
sets $S_{i}$ and $S_{j}$ by the nominator of Eq. (\ref{eq:stationary_pij_1})
and denoted it by $\mathbf{C}_{Eq}^{\tau}(i,j)$. Also, we have adopted
the usual notation $\pi_{i}=\int_{S_{i}}\mathrm{dx}\,\pi(x)$ for
the equilibrium probabilities of the discrete states. Such a matrix
of conditional transition probabilities is called a Markov state model
(MSM) or Markov model. It can be used as a simplified model for the
dynamics allowing extensive analysis, see Ref. \cite{PrinzEtAl_JCP10_MSM1}.

From a long simulation $X_{k},\,k=1,\ldots,K$ that samples points
from the stationary density $\pi$, the matrix $\mathbf{T}_{Eq}^{\tau}$
can be estimated by the formula
\begin{eqnarray}
\mathbf{T}_{Eq}^{\tau}(i,j) & \approx & \frac{\mathbf{S}^{\tau}(i,j)}{\mathbf{s}(i)}.\label{eq:Estimator_From_Count_0}
\end{eqnarray}

\subsection{Starting from local Equilibrium}

\label{subsec:LocalEq}

In practice, producing simulation data that samples from the global
equilibrium density $\pi$ is often not tractable. One of the strengths
of Markov models is the fact that the transition matrix can also be
expressed in terms of local equilibrium densities
\begin{eqnarray}
\pi_{S_{i}}(x) & = & \frac{1}{\pi_{i}}\chi_{i}(x)\pi(x).\label{eq:Local_Eq_Density}
\end{eqnarray}
The density $\pi_{S_{i}}$ is the normalized restriction of $\pi$
to state $S_{i}$. A Markov model transition matrix can also be estimated
by preparing an ensemble of trajectories in such a way that, within
each state, the distribution of starting points equals the local density
Eq. (\ref{eq:Local_Eq_Density}). These trajectories are simulated
for a very short time, and the fraction of trajectories starting in
$S_{i}$ and ending up in $S_{j}$ provides an estimate for the transition
matrix entry $\mathbf{T}_{Eq}^{\tau}(i,j)$ \cite{Weber:2006aa,Roeblitz_PhD}.
To see this, note that in the setting just described, the initial
distribution is a convex combination $\rho_{L}$ of the local densities
$\pi_{_{S_{i}}}$:
\begin{eqnarray}
\rho_{L} & = & \sum_{i=1}^{N}a_{i}\pi_{S_{i}},\label{eq:LocalEq_Dist_0}\\
\sum_{i=1}^{N}a_{i} & = & 1.\label{eq:LocalEq_Dist_1}
\end{eqnarray}
Here, $a_{i}$ is the probability to start in state $S_{i}$. Upon
replacing $\pi$ by $\rho_{L}$ in Eq. (\ref{eq:stationary_pij_1}),
it follows that
\begin{eqnarray}
\mathbf{T}_{Eq}^{\tau}(i,j) & = & \frac{\int_{S_{i}}\mathrm{dx}\int_{S_{j}}\mathrm{dy}\,\rho_{L}(x)p(x,y;\tau)}{\int_{S_{i}}\mathrm{dx}\,\rho_{L}(x)}.\label{eq:T_Eq_Local}
\end{eqnarray}
Only very short trajectories and knowledge of the local densities
are needed for the application of this method. However, this method
suffers from three major disadvantages: first, the intermediate data
points of the simulations cannot be used. Second, estimation of the
local densities requires the use of biased sampling methods, which
is a significant extra effort and entails additional difficulties.
Third, changing the discretization requires to redo the simulations,
which is not acceptable if a suitable discretization is not easy to
find.

\subsection{Multiple Step Estimator}

\label{subsec:multi-step-estimator}

A common way to construct MSMs in practice is by conducting a large
set of distributed simulations $\mathbf{Y}_{q},\,q=1,\ldots,Q$ of
lengths that are shorter than the largest relaxation timescales of
the system, but are longer than the lag time $\tau$. For our theoretical
investigation we will assume that each of these trajectories has the
same length of $K$ stored simulation steps, but for the estimators
we will be deriving later uniform length is not a requirement, see
Appendix \ref{sec:app_B_simulation_length}.

The simulations are started from some arbitrary initial distribution
at time $k=1$. The transition probability matrix is estimated by
replacing $\mathbf{S}(i,j)$ and $\mathbf{s}(i)$ by their empirical
mean values over all simulations $\mathbf{Y}_{q}$. These are defined
by the following equations, where we include the corresponding definition
for $\mathbf{S}_{r}^{2\tau}$ for later use:

\begin{eqnarray}
\overline{\mathbf{s}} & = & \frac{1}{Q}\sum_{q=1}^{Q}\mathbf{s}(\mathbf{Y}_{q}),\label{eq:Empirical_Mean_Si}\\
\overline{\mathbf{S}}^{\tau} & = & \frac{1}{Q}\sum_{q=1}^{Q}\mathbf{S}^{\tau}(\mathbf{Y}_{q}),\label{eq:Empirical_Mean_Stau}\\
\overline{\mathbf{S}}_{r}^{2\tau} & = & \frac{1}{Q}\sum_{q=1}^{Q}\mathbf{S}_{r}^{2\tau}(\mathbf{Y}_{q}).\label{eq:Empirical_Mean_S2tau}
\end{eqnarray}
In analogy to Eq. (\ref{eq:Estimator_From_Count_0}), the transition
matrix is then estimated by

\begin{eqnarray}
\overline{\mathbf{T}}^{\tau}(i,j) & = & \frac{\overline{\mathbf{S}}^{\tau}(i,j)}{\overline{\mathbf{s}}(i)}.\label{eq:Empirical_Mean_TransitionMatrix}
\end{eqnarray}

Additional constraints can be incorporated in order to obtain more
specific estimators than Eq. (\ref{eq:Empirical_Mean_TransitionMatrix}),
such as estimators obeying detailed balance \cite{Bowman_JCP09_Villin,PrinzEtAl_JCP10_MSM1,Trendelkamp-Schroer:2015aa}.

\begin{figure}
\begin{centering}
\includegraphics[height=0.5\paperheight]{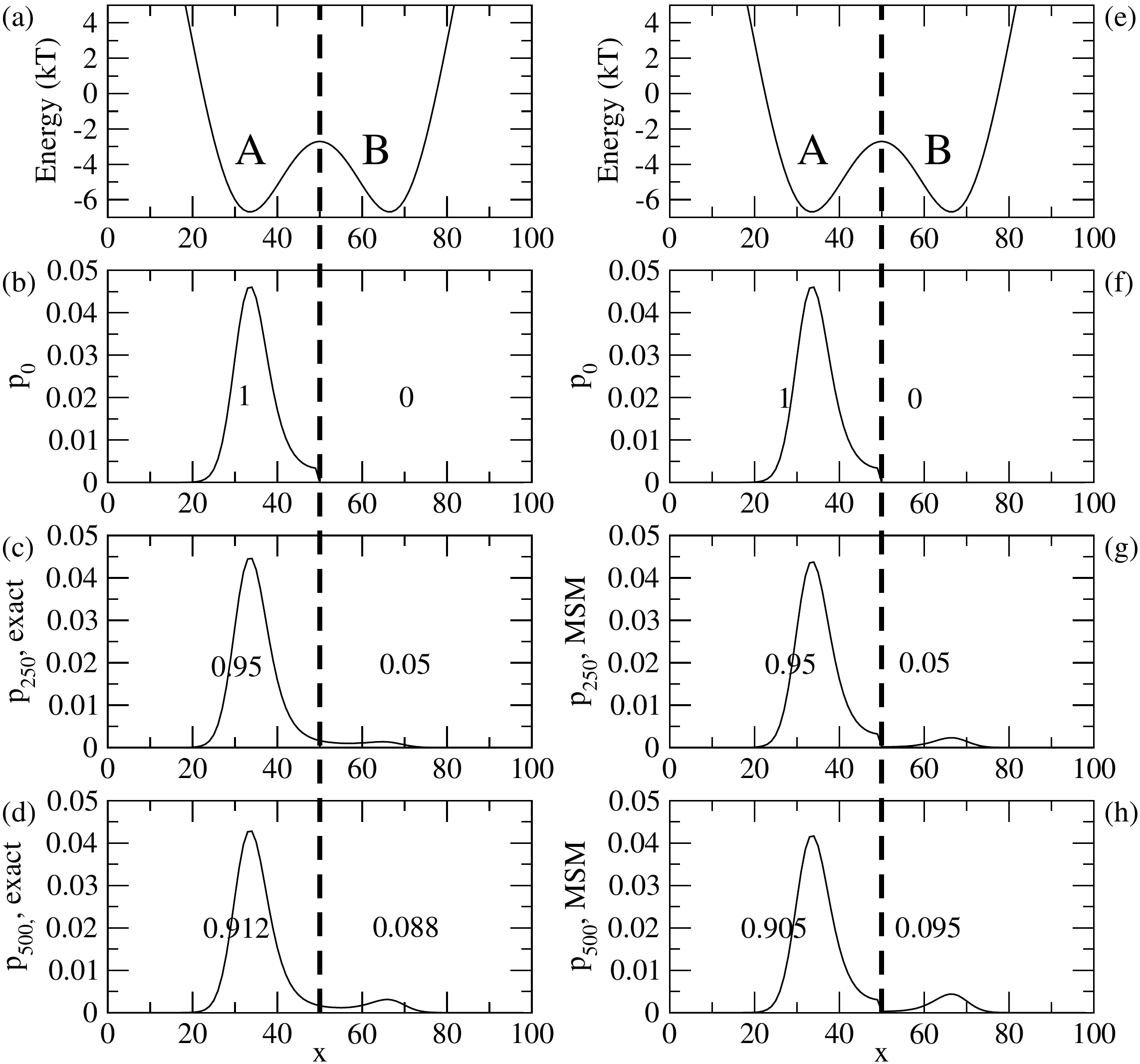}
\par\end{centering}
\caption{Loss of local equilibrium property illustrated by comparing the dynamics
of the diffusion in a double-well potential (a,e) at time steps 0
(b), 250 (c), 500 (d) with the predictions of a Markov model parameterized
at lag time $\tau=250$ at the same times 0 (f), 250 (g), 500 (h).
Please refer to the supplementary material of Ref. \cite{PrinzEtAl_JCP10_MSM1}
for details of the system. (b, c, d) show the true distribution of
the system (solid black line) and the probabilities associated with
the two discrete states left and right of the dashed line. The numbers
in (f, g, h) are the discrete state probabilities $p_{i}(k\tau),\,i=1,2,\,k=0,1,2$,
predicted by the Markov model. The solid black lines shows the hypothetical
density $p_{i}(k\tau)\pi_{S_{i}}$ that is inherently assumed when
estimating a Markov model by counting transitions over multiple steps.
This figure has been re-used with permission from Ref. \cite[Fig. 4]{PrinzEtAl_JCP10_MSM1},
copyright 2011, American Institute of Physics.\label{fig:fig2-illustration_localeq}}
\end{figure}

The argument from Sec. \ref{subsec:LocalEq} cannot be transferred
directly to a multiple step estimator like Eqs. (\ref{eq:Empirical_Mean_Si}-\ref{eq:Empirical_Mean_Stau}):
Even if the simulations are started from local equilibrium, this property
is lost after the first simulation step, and the resulting estimates
are no longer unbiased. A detailed illustration of this phenomenon
has been provided by Ref. \cite[Fig. 4]{PrinzEtAl_JCP10_MSM1}, and
we repeat it here in Figure \ref{fig:fig2-illustration_localeq}.
It can be argued that if the discretization is chosen well enough
such that the dynamics equilibrates to an approximate local equilibrium
within all states over a single time step, the bias can be expected
to be very small. This assumption is difficult to check or quantify
in practice. In the next section, we analyze the bias introduced by
the multiple step estimator, as well as its dependence on the lag
time and simulation length.

\subsection{Estimation Error from Non-Equilibrium Simulations}

Now we study the effect of using an initial distribution of simulation
data that is not in local equilibrium when the transitions are counted.
This deviation from local equilibrium could come either from the fact
that we start trajectories in an arbitrary initial condition, or that
our trajectories exceed the lag time $\tau$ such that an initially
prepared local equilibrium is lost for all transition counts harvested
after the first one (Sec. \ref{subsec:multi-step-estimator}).

Let $\rho$ denote the \textit{empirical }distribution sampled by
the simulations. We need to study the error between the equilibrium
transition matrix $\mathbf{T}_{Eq}^{\tau}$ and the asymptotic limit
of Eq. (\ref{eq:Empirical_Mean_TransitionMatrix}). To this end, we
study the asymptotic limits of $\overline{\mathbf{S}}^{\tau}(i,j)$
and $\overline{\mathbf{s}}(i)$ in the limit of infinitely many simulations,
$Q\rightarrow\infty$, but each having finite lengths:

\begin{eqnarray}
\mathbf{C}_{\rho}^{\tau}(i,j) & := & \mathbb{E}\left(\mathbf{S}^{\tau}(i,j)\right),\label{eq:Definition_Ctau}\\
\mathbf{c}_{\rho}(i) & := & \mathbb{E}\left(\mathbf{s}(i)\right),\label{eq:Definition_ci}\\
\mathbf{T}_{\rho}^{\tau}(i,j) & := & \frac{\mathbf{C}_{\rho}^{\tau}(i,j)}{\mathbf{c}_{\rho}(i)}.\label{eq:Expected_Ttau}
\end{eqnarray}

Thus, we use the symbols $\mathbf{C}_{\rho}^{\tau},\,\mathbf{c}_{\rho}$
for the \textit{expected} count matrix and vector of total counts
associated with the empirical distribution $\rho$. Using the spectral
decomposition Eq. (\ref{eq:SpectralDecompKernel}), the expected count
matrix can be expressed in terms of the spectral components of the
dynamics:
\begin{eqnarray}
\mathbf{C}_{\rho}^{\tau}(i,j) & = & \sum_{m=1}^{M}\lambda_{m}(\tau)\langle\chi_{i},\psi_{m}\rangle_{\rho}\langle\chi_{j},\psi_{m}\rangle_{\pi},\label{eq:Expectation_C^tau_0}\\
\langle\chi_{i},\psi_{m}\rangle_{\rho} & = & \int_{\Omega}\mathrm{dx}\,\chi_{i}(x)\psi_{m}(x)\rho(x),\label{eq:Expectation_C^tau_1}\\
\langle\chi_{i},\psi_{m}\rangle_{\pi} & = & \int_{\Omega}\mathrm{dx}\,\chi_{i}(x)\psi_{m}(x)\pi(x).\label{eq:Expectation_C^tau_2}
\end{eqnarray}
In matrix form, Eq. (\ref{eq:Expectation_C^tau_0}) can be written
as
\begin{eqnarray}
\mathbf{C}_{\rho}^{\tau} & = & \mathbf{Q}_{\rho}\boldsymbol{\Lambda}(\tau)\mathbf{Q}_{\pi}^{T},\label{eq:decomp_Ctau_0}\\
\mathbf{Q}_{\rho}(i,m) & = & \langle\chi_{i},\psi_{m}\rangle_{\rho},\label{eq:decomp_Ctau_1}\\
\mathbf{Q}_{\pi}(j,m) & = & \langle\chi_{j},\psi_{m}\rangle_{\pi}.\label{eq:decomp_Ctau_2}
\end{eqnarray}
These matrices contain the MSM projections of the true eigenfunctions,
i.e. their approximations by step functions, that is extensively discussed
in \cite{SarichNoeSchuette_MMS09_MSMerror,PrinzEtAl_JCP10_MSM1}.
Let us emphasize that Eq. (\ref{eq:Expectation_C^tau_0}) also holds
for arbitrary basis functions, i.e. $\chi_{i}$ is not required to
be a basis of indicator functions. Thus, it is the most general expression
for a correlation matrix from Markovian dynamics.

Summation over $j$ shows that
\begin{eqnarray}
\mathbf{c}_{\rho}(i) & = & \langle\chi_{i}\rangle_{\rho}.\label{eq:Expectation_c_i_0}
\end{eqnarray}
It follows from Eq. (\ref{eq:Expectation_C^tau_0}) that the spectral
expansion of $\mathbf{C}_{Eq}^{\tau}$ is given by
\begin{eqnarray}
\mathbf{C}_{Eq}^{\tau}(i,j) & = & \sum_{m=1}^{M}\lambda_{m}(\tau)\langle\chi_{i},\psi_{m}\rangle_{\pi}\langle\chi_{j},\psi_{m}\rangle_{\pi},\label{eq:Spectral_Decomp_Eq_0}
\end{eqnarray}
using the fact that for trajectories started from global equilibrium
we have $\rho=\pi$. Combining Eqs. (\ref{eq:Expectation_C^tau_0}),
(\ref{eq:Expectation_c_i_0}) and (\ref{eq:Spectral_Decomp_Eq_0}),
we obtain an expression for the estimation error $\mathbf{E}^{\tau}:=\mathbf{T}_{\rho}^{\tau}-\mathbf{T}_{Eq}^{\tau}$:

\begin{eqnarray}
\mathbf{E}^{\tau}(i,j) & = & \frac{\mathbf{C}_{\rho}^{\tau}(i,j)}{\mathbf{c}_{\rho}(i)}-\frac{\mathbf{C}_{Eq}^{\tau}(i,j)}{\pi_{i}}\label{eq:Estimation_Error_0}\\
 & = & \sum_{m=2}^{M}\lambda_{m}(\tau)\langle\chi_{j},\psi_{m}\rangle_{\pi}\left[\frac{\langle\chi_{i},\psi_{m}\rangle_{\rho}}{\langle\chi_{i}\rangle_{\rho}}-\frac{\langle\chi_{i},\psi_{m}\rangle_{\pi}}{\langle\chi_{i}\rangle_{\pi}}\right]\label{eq:Estimation_Error_1}\\
 & = & \sum_{m=2}^{M}\lambda_{m}(\tau)\langle\chi_{j},\psi_{m}\rangle_{\pi}\left[\frac{\langle\chi_{i},\psi_{m}-q_{im}\psi_{1}\rangle_{\rho}}{\langle\chi_{i}\rangle_{\rho}}\right],\label{eq:Estimation_Error_2}
\end{eqnarray}
where $q_{im}=\frac{\langle\chi_{i},\psi_{m}\rangle_{\pi}}{\langle\chi_{i}\rangle_{\pi}}$,
and we were able to drop the $m=1$ terms on both sides as they are
equal. Inspecting this expression leads to a number of insights that
are practically important for analyzing simulation data with MSMs:
\begin{enumerate}
\item \textbf{MSM estimation from long trajectories}: In the limit that
our trajectories are longer than the timescale of the slowest process,
the empirical distribution $\rho$ converges to the equilibrium distribution
$\pi$, and the bias becomes zero. This offers an explanation why
MSMs built from ultra-long simulations \cite{Shaw_Science10_Anton,Lindorff-Larsen:2011aa}
are quite well-behaved and have been extensively used for benchmarking
and method validation.
\item \textbf{MSM estimation from short trajectories}: Even if the trajectories
are not long enough to reach global equilibrium, because of Eq. (\ref{eq:Def_timescales}),
the bias decays multi-exponentially with the lag time $\tau$. This
is an important insight, because MSMs are in practice constructed
in the limit of long enough lag times in which the timescale estimates
converge \cite{Swope:2004aa,PrinzEtAl_JCP10_MSM1}, and the above
equation shows that this limit is meaningful as it approaches an unbiased
estimate. 
\item \textbf{Dependence of bias on the discretization error}: The above
formula reflects the well-known insight that Markov models are free
of bias if the discretization perfectly approximates the dominant
eigenfunctions, meaning that the eigenfunctions are constant on the
states $S_{i}$ \cite{Swope:2004aa,PrinzEtAl_JCP10_MSM1}.
\item \textbf{Consequences for adaptive sampling}: Previous adaptive sampling
approaches have suggested to prepare an initial local equilibrium
distribution in order to shoot trajectories out of selected states
\cite{Roeblitz_PhD}. The above analysis shows that this strategy
is effective if we only count a single transition out of the state,
but is ineffective when longer trajectories are shot. In the latter
case, it is simpler to ignore the initial distribution and to reduce
the effect of bias by extending the lag time $\tau$, see again Fig.
\ref{fig:intro_figure} and also the next example.
\end{enumerate}

\subsection{Example}

\begin{figure}
\begin{centering}
\includegraphics{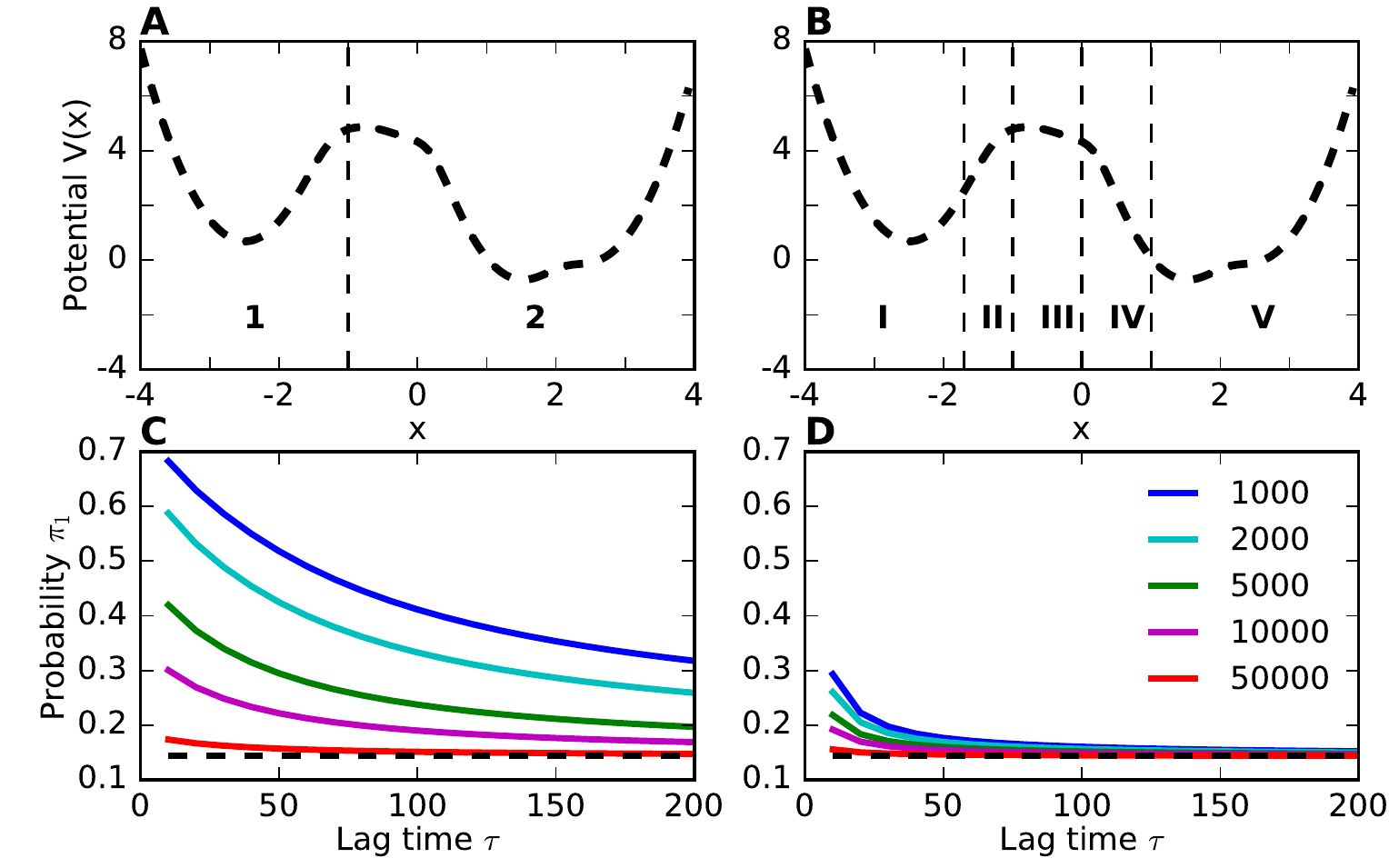}
\par\end{centering}
\caption{A, B: One-dimensional potential function with two different discretizations
into two states and five states, resp. C: Expected estimate of the
equilibrium probability of state 1 as a function of the lag time,
for simulation lengths $K=1000,\,2000,\,5000,\,10000,\,50000$, and
using the discretization from panel A. The simulations are initiated
in local equilibrium in both states 1 and 2, but predominantly in
state 1 ($a_{1}=0.9,\,a_{2}=0.1$). D: The same for the five state
discretization from panel B.\label{fig:fig3-Illustration}}
\end{figure}

Before proceeding, we illustrate these findings by re-visiting the
one-dimensional model system presented in the introduction. We study
the same two different discretizations, the two state model from panel
A of Fig. \ref{fig:fig3-Illustration} and the five state discretization
shown in Fig. \ref{fig:fig3-Illustration} B. Again, simulations are
initiated from local equilibrium in states 1 and 2 of the coarse discretization,
with $a_{1}=0.9,\,a_{2}=0.1$. We study the expected estimate of the
equilibrium probability of state 1, which equals the equilibrium probability
of states I and II for the finer state definition. Panels C and D
of Fig. \ref{fig:fig3-Illustration} show the respective estimates
for the coarse and fine discretization as a function of the lag time,
for simulation lengths $K=1000,\,2000,\,5000,\,10000,\,50000$. Indeed,
the estimates improve if the lag time is increased, if the simulation
length is increased, or if the discretization is improved. From the
coarse partitioning example, we conclude that relaxation to global
equilibrium can be required in order to obtain unbiased estimates
from simulations initiated out of local equilibrium.

\section{Correction of Estimation Bias using Observable Operator Models}

\label{sec:OOMs}

In this section, we show how to go beyond just using a longer lag
time $\tau$ and suggest correction mechanisms to obtain the correct
equilibrium transition matrix $\mathbf{T}_{Eq}^{\tau}$ (Eqs. (\ref{eq:stationary_pij_0}-\ref{eq:stationary_pij_2}))
from an ensemble of short simulations. This can be accomplished regardless
of the starting distribution being in global equilibrium, in local
equilibrium, or far from any equilibrium.

As discussed above, limitations of MSMs include the assumption of
Markovianity, sensitivity to projection error, and sensitivity to
the distribution of trajectory starting points. All of these limitations
can be overcome by realizing that molecular dynamics that is observed
in a chosen set of variables, reaction coordinates or order parameters
at a certain lag time $\tau$ can be \emph{exactly} described by projected
Markov models (PMMs) \cite{Noe:2013aa}. This insight allows us to
employ estimators that are not affected by the MSM limitations, such
as hidden Markov models (HMMs) \cite{Noe:2013aa} or observable operator
models (OOMs) \cite{Jaeger:2000aa,Wu:2015aa,Wu:2016aa}, that operate
on the discretized state space. 

Here, we employ OOMs in order to get improved MSM estimators that
are not subject to the bias caused by a non-equilibrium distribution
of the trajectories used. In a nutshell, OOMs are spectral estimators
able to provide unbiased estimates of stationary and dynamical quantities
for dynamical systems that can be well described by a finite number
of dynamical components. Here we only summarize a few aspects of OOMs
that are relevant to the present paper and present an algorithm that
can be used to estimate MSMs without bias from the initial trajectory
distribution. To fully understand the theoretical background and derivation,
please refer to \cite{Jaeger:2000aa,Wu:2015aa,Wu:2016aa}.

\subsection{Observable Operator Models}

\textit{Observable operator models} (OOMs) provide a framework that
completely captures the dynamics of a stochastic dynamical system
by a finite-dimensional algebraic system if only a finite number $M$
of relaxation processes contribute in Eq. (\ref{eq:SpectralDecompKernel}),
see Refs. \cite{Jaeger:2000aa,Wu:2015aa}. For molecular dynamics,
this property is achieved if we observe and model the dynamics at
a finite lag time $\tau$. The \textit{full-state observable operator
}$\boldsymbol{\Xi}_{\Omega}$ is an $M\times M$ matrix which contains
the scalar products between the eigenfunctions: 
\begin{eqnarray}
\boldsymbol{\Xi}_{\Omega}(m,m') & = & \lambda_{m}(\tau)\int_{\Omega}\mathrm{dx}\,\psi_{m}(x)\psi_{m'}(x)\pi(x).\label{eq:Definition_Full_Obs_Operator}
\end{eqnarray}
In statistical terms, $\boldsymbol{\Xi}_{\Omega}$ is the expectation
value of the covariance matrix between eigenfunctions. As eigenfunctions
are orthogonal with respect to the equilibrium distribution $\pi$,
or in other words, statistically uncorrelated, $\boldsymbol{\Xi}_{\Omega}$
is just a diagonal matrix of the eigenvalues:
\begin{eqnarray}
\boldsymbol{\Xi}_{\Omega} & = & \boldsymbol{\Lambda}.\label{eq:Full_State_Obs_Op}
\end{eqnarray}

If we do not integrate over the full state space $\Omega$ in Eq.
(\ref{eq:Definition_Full_Obs_Operator}), but only over a subset $A\subset\Omega$,
we can define a matrix $\boldsymbol{\Xi}_{A}$ of size $M\times M$,
called the \textit{set-observable operator }for set $A$\textit{.}
All set-observable operators and two vectors $\boldsymbol{\omega},\,\boldsymbol{\sigma}\in\mathbb{R}^{M}$
are the key ingredients of OOM theory. The vectors $\boldsymbol{\omega},\,\boldsymbol{\sigma}$
equal the first canonical unit vector $\mathbf{e}_{1}$, i.e. $\boldsymbol{\omega}=\boldsymbol{\sigma}=\mathbf{e}_{1}=\left(1,0,\ldots,0\right)^{T}$,
and they are called \textit{information state} and \textit{evaluator},
respectively. If the finite-rank assumption Eq. (\ref{eq:SpectralDecompKernel})
holds, these components form an algebraic system that allows to compute
equilibrium probabilites of finite observation sequences. Let $A_{1},\ldots,A_{l}$
be arbitrary subsets of $\Omega$ that do not need to form a partition
of the state space. If Eq. (\ref{eq:SpectralDecompKernel}) is satisfied,
we can compute the probability that a trajectory in equilibrium visits
set $A_{1}$ at time $\tau$, set $A_{2}$ at time $2\tau$, ...,
and set $A_{l}$ at time $l\tau$ by the following matrix-vector product:

\begin{eqnarray}
\mathbb{P}(X_{\tau}\in A_{1},X_{2\tau}\in A_{2},\ldots,X_{l\tau}\in A_{l}) & = & \boldsymbol{\omega}^{T}\boldsymbol{\Xi}_{A_{1}}\ldots\boldsymbol{\Xi}_{A_{l}}\boldsymbol{\sigma}.\label{eq:Path_Prob_OOM}
\end{eqnarray}

The proof can be found in Ref. \cite{Wu:2015aa}, we also repeat it
in Appendix \ref{sec:Appendix_OOM_Path_Prob}. Note that, in case
that $A_{1},\ldots,A_{l}$ form a partition of state space, the probability
of such an observation sequence cannot be computed from a Markov model
transition matrix between the sets $A_{1},\ldots,A_{l}$, unless the
dynamics is Markovian on these sets. This clearly distinguishes an
OOM from a Markov model: An OOM can correctly describe arbitrary projected
dynamics as long as Eq. (\ref{eq:SpectralDecompKernel}) holds.

As a Markov process is determined entirely by finite observation probabilities
like Eq. (\ref{eq:Path_Prob_OOM}), it follows that we can compute
several key equilibrium, kinetic and mechanistic quantities in an
unbiased fashion if we can somehow estimate the OOM components. For
a fixed decomposition of state space into sets $S_{r},\,r=1,\ldots,N$
as before, let us denote the set-observable operators of sets $S_{r}$
by $\boldsymbol{\Xi}_{r}$, which implies that

\begin{eqnarray}
\boldsymbol{\Xi}_{\Omega} & = & \sum_{r=1}^{N}\boldsymbol{\Xi}_{r}.\label{eq:Sum_Set_Obs_Operators}
\end{eqnarray}

It follows from Eq. (\ref{eq:Path_Prob_OOM}) that we can compute
the unbiased equilibrium correlation matrix and the stationary probabilities
by the formulas
\begin{eqnarray}
\mathbf{C}_{Eq}^{\tau}(i,j) & = & \boldsymbol{\omega}^{T}\boldsymbol{\Xi}_{i}\boldsymbol{\Xi}_{j}\boldsymbol{\sigma},\label{eq:OOM_Path_Prob_C_Eq}\\
\pi_{i} & = & \boldsymbol{\omega}^{T}\boldsymbol{\Xi}_{i}\boldsymbol{\sigma}.\label{eq:OOM_Path_Prob_pi_i}
\end{eqnarray}

In practice we cannot directly estimate $\boldsymbol{\Xi}_{r}$ but
only a \emph{similar} operator $\hat{\boldsymbol{\Xi}}_{r}$. However,
it follows directly from Eqs. (\ref{eq:OOM_Path_Prob_C_Eq}-\ref{eq:OOM_Path_Prob_pi_i})
that if an unknown similarity transform $\mathbf{R}\in\mathbb{R}^{M\times M}$
affects all OOM quantities via
\begin{eqnarray}
\hat{\boldsymbol{\Xi}}_{r} & = & \mathbf{R}\boldsymbol{\Xi}_{r}\mathbf{R}^{-1},\label{eq:Equiv_OOM_0}\\
\hat{\boldsymbol{\omega}}^{T} & = & \boldsymbol{\omega}^{T}\mathbf{R}^{-1},\label{eq:Equiv_OOM_1}\\
\hat{\boldsymbol{\sigma}} & = & \mathbf{R}\boldsymbol{\sigma},\label{eq:Equiv_OOM_2}
\end{eqnarray}
then Eqs. (\ref{eq:OOM_Path_Prob_C_Eq}-\ref{eq:OOM_Path_Prob_pi_i})
remain exactly valid using $\hat{\boldsymbol{\omega}},\,\hat{\boldsymbol{\Xi}}_{r},\,\hat{\boldsymbol{\sigma}}$.
In other words, all OOMs that can be constructed by choosing some
transformation matrix $\mathbf{R}$ form a family of equivalent OOMs.
A specific member of this family \emph{can} be estimated directly
from simulation data, and thus we can use it in order to obtain unbiased
estimates of Eqs. (\ref{eq:OOM_Path_Prob_C_Eq}-\ref{eq:OOM_Path_Prob_pi_i})
even from a large ensemble of trajectories that do not need to sample
from global equilibrium. It has been shown in Ref. \cite{Wu:2016aa}
that Eqs. (\ref{eq:Estimator_E(S_i)}-\ref{eq:Estimator_sigma}) and
(\ref{eq:Estimator_omega0_0}-\ref{eq:Estimator_omega0_1}) in the
next subsection indeed provide the components of an equivalent OOM,
i.e. there is an invertible matrix $\mathbf{R}$ s.t. Eqs. (\ref{eq:Equiv_OOM_0}-\ref{eq:Equiv_OOM_2})
are satisfied in the absence of statistical noise. 

\subsection{Unbiased Estimation of Markov State Models}

\label{subsec:Unbiased_MSM_Estimator}

To construct an \emph{exact} unbiased estimator we need three ingredients:
(i) the expectation values of the empirical count matrix $\mathbf{C}_{\rho}^{\tau}$,
(ii) the vector of total counts $\mathbf{c}_{\rho}$ from Eqs. (\ref{eq:Definition_Ctau}-\ref{eq:Definition_ci}),
and additionally (iii) the two-step count matrices 

\begin{eqnarray}
\mathbf{C}_{\rho,r}^{2\tau} & := & \mathbb{E}\left(\mathbf{S}_{r}^{2\tau}\right).\label{eq:Definition_C2tau}
\end{eqnarray}

As a reminder, expectation values here denote the expectation over
a trajectory ensemble sampling from the empirical (non-equilibrium)
distribution $\rho$. In practice, only finitely many simulations
are available, and we thus replace $\mathbf{c}_{\rho},\,\mathbf{C}_{\rho}^{\tau}$
and $\mathbf{C}_{\rho,r}^{2\tau}$ by count vectors and matrices $\overline{\mathbf{s}},\,\overline{\mathbf{S}}^{\tau}$
and $\overline{\mathbf{S}}_{r}^{2\tau}$ (Eqs. (\ref{eq:Empirical_Mean_Si}-\ref{eq:Empirical_Mean_S2tau})),
which are asymptotically unbiased estimators. The unbiased estimation
algorithm can be summarized as follows:
\begin{enumerate}
\item Obtain the empirical mean $\overline{\mathbf{s}}$, count matrix $\overline{\mathbf{S}}^{\tau}$
and two-step count matrices $\overline{\mathbf{S}}_{r}^{2\tau}$ from
simulation data using Eqs. (\ref{eq:Empirical_Mean_Si}-\ref{eq:Empirical_Mean_S2tau}).
\item Decompose the count matrix $\overline{\mathbf{S}}^{\tau}$ by singular
value decomposition (SVD)
\begin{eqnarray}
\overline{\mathbf{S}}^{\tau} & = & \mathbf{V}\Sigma\mathbf{W}^{T},\label{eq:SVD_Ctau}
\end{eqnarray}
and compute weighted projections onto the leading $M$ left and right
singular vectors by
\begin{eqnarray}
\mathbf{F}_{1} & = & \mathbf{V}_{M}\Sigma_{M}^{-1/2},\label{eq:Definition_F1_F2_0}\\
\mathbf{F}_{2} & = & \mathbf{W}_{M}\Sigma_{M}^{-1/2}.\label{eq:Definition_F1_F2_1}
\end{eqnarray}
We have used the symbols $\mathbf{V}_{M},\,\mathbf{W}_{M},\,\Sigma_{M}$
to denote the restriction of these matrices to their first $M$ columns. 
\item Use $\mathbf{F}_{1},\,\mathbf{F}_{2}$ to obtain the set-observable
operators $\hat{\boldsymbol{\Xi}}_{r}$ and the evaluation state vector
$\hat{\boldsymbol{\sigma}}$ of an equivalent OOM via
\begin{eqnarray}
\hat{\boldsymbol{\Xi}}_{r} & = & \mathbf{F}_{1}^{T}\overline{\mathbf{S}}_{r}^{2\tau}\mathbf{F}_{2},\label{eq:Estimator_E(S_i)}\\
\hat{\boldsymbol{\sigma}} & = & \mathbf{F}_{1}^{T}\overline{\mathbf{s}}.\label{eq:Estimator_sigma}
\end{eqnarray}
Compute the full-state observable operator $\hat{\boldsymbol{\Xi}}_{\Omega}=\sum_{r=1}^{N}\hat{\boldsymbol{\Xi}}_{r}$
and obtain the information state vector $\hat{\boldsymbol{\omega}}$
as the solution to the eigenvalue problem:
\begin{eqnarray}
\hat{\boldsymbol{\omega}}^{T}\hat{\boldsymbol{\Xi}}_{\Omega} & = & \hat{\boldsymbol{\omega}}^{T},\label{eq:Estimator_omega0_0}\\
\hat{\boldsymbol{\omega}}^{T}\hat{\boldsymbol{\sigma}} & = & 1.\label{eq:Estimator_omega0_1}
\end{eqnarray}
The normalization Eq. (\ref{eq:Estimator_omega0_1}) can be achieved
by dividing the arbitrarily scaled solution $\hat{\boldsymbol{\omega}}^{T}$
by $\hat{\boldsymbol{\omega}}^{T}\hat{\boldsymbol{\sigma}}$.
\item Compute the unbiased equilibrium correlation matrix and unbiased equilibrium
distribution by
\begin{eqnarray}
\mathbf{C}_{Eq}^{\tau}(i,j) & = & \hat{\boldsymbol{\omega}}^{T}\hat{\boldsymbol{\Xi}}_{i}\hat{\boldsymbol{\Xi}}_{j}\hat{\boldsymbol{\sigma}},\label{eq:OOM_Estimator_Ctau_Eq}\\
\pi_{i} & = & \hat{\boldsymbol{\omega}}^{T}\hat{\boldsymbol{\Xi}}_{i}\hat{\boldsymbol{\sigma}}\label{eq:OOM_Estimator_pi_0}\\
 & = & \sum_{j=1}^{N}\mathbf{C}_{Eq}^{\tau}(i,j).\label{eq:OOM_Estimator_pi_1}
\end{eqnarray}
and then obtain the unbiased MSM transition matrix $\mathbf{T}_{Eq}^{\tau}$
either using the nonreversible estimator
\begin{equation}
\mathbf{T}_{Eq}^{\tau}(i,j)=\frac{\mathbf{C}_{Eq}^{\tau}(i,j)}{\pi_{i}},\label{eq:OOM_Estimator_Ttau_Eq}
\end{equation}
or the reversible estimator
\begin{equation}
\mathbf{T}_{Eq}^{\tau}(i,j)=\frac{\mathbf{C}_{Eq}^{\tau}(i,j)+\mathbf{C}_{Eq}^{\tau}(j,i)}{\sum_{j=1}^{N}\mathbf{C}_{Eq}^{\tau}(i,j)+\sum_{j=1}^{N}\mathbf{C}_{Eq}^{\tau}(j,i)}.\label{eq:OOM_Estimator_Ttau_Eq_Rev}
\end{equation}
\end{enumerate}
Let us briefly comment on the central idea behind this algorithm,
which is the estimation of an equivalent OOM in the third step, particularly
in Eq. (\ref{eq:Estimator_E(S_i)}). Using the path probability formula
Eq. (\ref{eq:Path_Prob_OOM}), it can be shown that the expected two-step
count matrix is given by

\begin{eqnarray}
\mathbf{C}_{\rho,r}^{2\tau} & = & \mathbf{Q}_{\rho}\boldsymbol{\Xi}_{r}\boldsymbol{\Lambda}(\tau)\mathbf{Q}_{\pi}^{T},\label{eq:decomp_C^2tau_r}
\end{eqnarray}
where the matrices $\mathbf{Q}_{\rho},\,\mathbf{Q}_{\pi}$ are the
same as in Eqs. (\ref{eq:decomp_Ctau_1}-\ref{eq:decomp_Ctau_2}).
Thus, by the intermediate step, the set-observable operator is introduced
into the decomposition of the two-step count matrix. Now, the idea
is to find two matrices $\mathbf{F}_{1},\,\mathbf{F}_{2}\in\mathbb{R}^{N\times M}$,
such that $\mathbf{R}_{1}:=\mathbf{F}_{1}^{T}\mathbf{Q}_{\rho}$ and
\textbf{$\mathbf{R}_{2}:=\boldsymbol{\Lambda}(\tau)\mathbf{Q}_{\pi}^{T}\mathbf{F}_{2}$}
are inverse to each other, because this implies that

\begin{eqnarray}
\mathbf{F}_{1}^{T}\mathbf{C}_{\rho,r}^{2\tau}\mathbf{F}_{2} & = & \mathbf{R}_{1}\boldsymbol{\Xi}_{r}\mathbf{R}_{2}\label{eq:transform_C2tau_0}\\
 & = & \mathbf{R}\boldsymbol{\Xi}_{r}\mathbf{R}^{-1}\label{eq:transform_C2tau_1}
\end{eqnarray}
is the $r$-th component of an equivalent OOM. The properties of SVD
and the decomposition Eq. (\ref{eq:decomp_Ctau_0}) guarantee that
the choice of $\mathbf{F}_{1},\,\mathbf{F}_{2}$ in the second step
above achieves this goal:

\begin{eqnarray}
\mathbf{Id} & = & \mathbf{F}_{1}^{T}\mathbf{C}_{\rho}^{\tau}\mathbf{F}_{2}\label{eq:transform_Ctau_0}\\
 & = & \left(\mathbf{F}_{1}^{T}\mathbf{Q}_{\rho}\right)\left(\boldsymbol{\Lambda}(\tau)\mathbf{Q}_{\pi}^{T}\mathbf{F}_{2}\right)\label{eq:transform_Ctau_1}\\
 & = & \mathbf{R}_{1}\mathbf{R}_{2}.\label{eq:transform_Ctau_2}
\end{eqnarray}
Similar arguments can be used to justify the equations for $\boldsymbol{\omega},\,\boldsymbol{\sigma}$.
We also note that different choices of $\mathbf{F}_{1},\,\mathbf{F}_{2}$
in step 2 are possible. For detailed explanations and proofs, please
refer to the previous publications \cite{Jaeger:2000aa,Wu:2015aa,Wu:2016aa}.

\subsection{Recovery of Exact Relaxation Timescales}

A remarkable by-product of the procedure described above is that the
transformed full-state two-step count matrix $\hat{\boldsymbol{\Xi}}_{\Omega}$
is similar to a diagonal matrix of the system eigenvalues $\lambda_{m}(\tau)$
\emph{without any MSM projection error.}\textit{\emph{ This has been
shown for equilibrium data in Ref. \cite{Prinz:2012aa} and also applies
to non-equilibrium data}} \cite{Wu:2015aa}:

\begin{eqnarray}
\hat{\boldsymbol{\Xi}}_{\Omega} & = & \mathbf{R}\boldsymbol{\Xi}_{\Omega}\mathbf{R}^{-1}\label{eq:EV_Decomp_E_Omega_0}\\
 & = & \mathbf{R}\boldsymbol{\Lambda}(\tau)\mathbf{R}^{-1}.\label{eq:EV_Decomp_E_Omega_1}
\end{eqnarray}
Thus, diagonalization of $\hat{\boldsymbol{\Xi}}_{\Omega}$ provides
an estimate of the leading system eigenvalues, and consequently also
of the relaxation rates or timescales, that is not distorted by the
fact that we coarse-grain the dynamics to a Markov chain between coarse
sets in state space. These eigenvalue and timescale estimates are
only subject to statistical error, but not to any MSM model error.
It is impossible to directly build an MSM that produces these timescales
- when an MSM is desired, the timescales can only be approximated,
and they will only be correct in the limit of long lag times and good
discretization.

However, the fact that we can get a model-free estimate of the eigenvalues
and relaxation timescales can be used to assess the discretization
quality: According to the variational principle of conformation dynamics
\cite{NoeNueske_MMS13_VariationalApproach}, the exact system eigenvalues
provide an upper bound to the eigenvalues of the equilibrium transition
matrix $\mathbf{T}_{Eq}^{\tau}$. By comparing the eigenvalues of
$\mathbf{T}_{Eq}^{\tau}$ to those from Eqs. (\ref{eq:EV_Decomp_E_Omega_0}-\ref{eq:EV_Decomp_E_Omega_1}),
the MSM discretization error theoretically studied in \cite{SarichNoeSchuette_MMS09_MSMerror,PrinzEtAl_JCP10_MSM1,Djurdjevac:2012aa}
can be practically quantified.

\subsection{Selection of Model Rank}

\label{subsec:Numerical_Rank}

\begin{figure}
\begin{centering}
\includegraphics[width=1\textwidth]{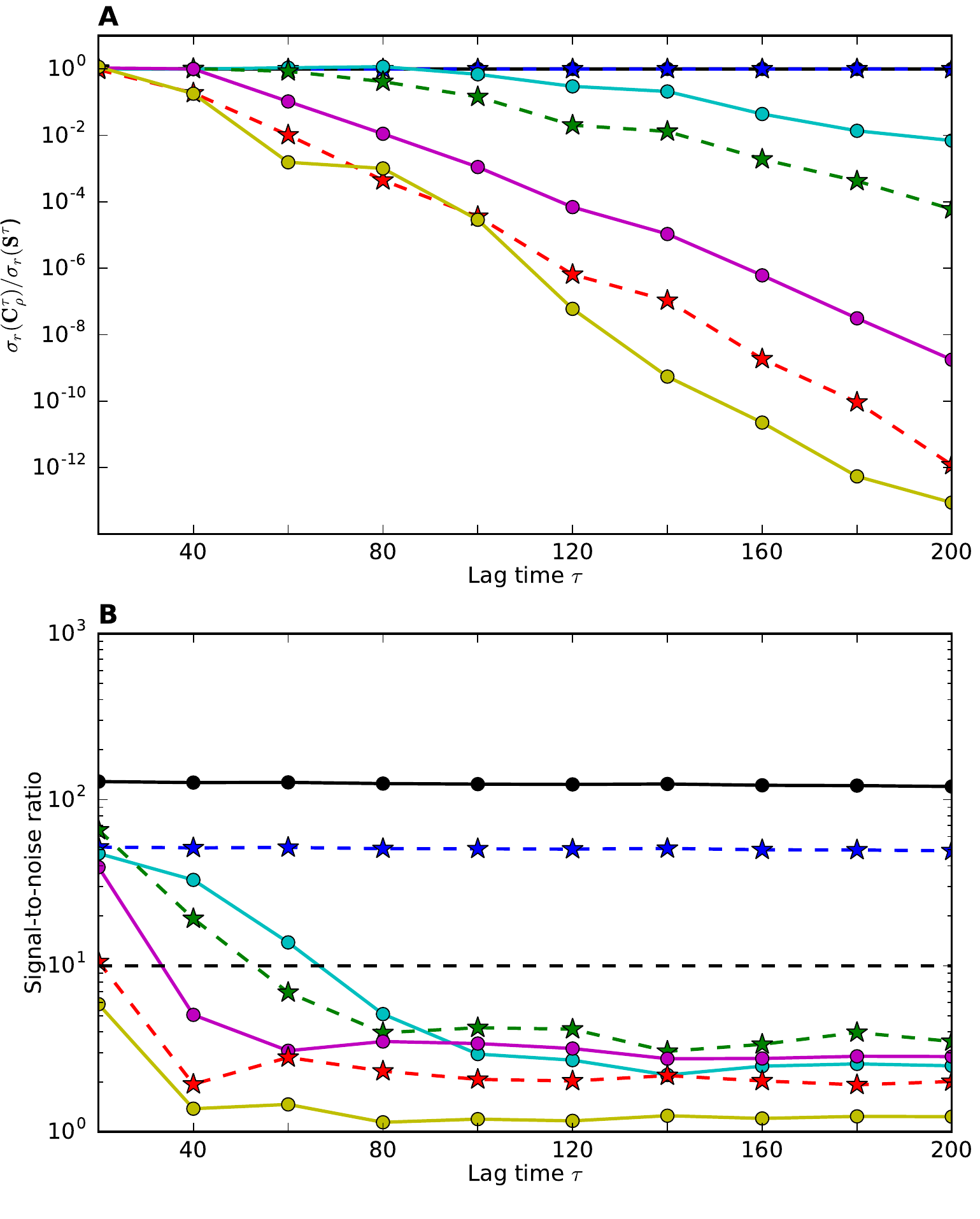}
\par\end{centering}
\caption{Analysis of statistical uncertainties for singular values of the count
matrix. We use the one-dimensional model system and seven state discretization
as in Sec. \ref{subsec:One_d_model_system}, the sample consists of
$Q=5000$ trajectories of length $K=2000$. A: For each of the seven
singular values (distinguished in descending order by the colors black,
blue, cyan, green, magenta, red and yellow), we show the ratio of
the true singular value $\sigma_{r}(\mathbf{C}_{\rho}^{\tau}),\,r=1,\ldots,7$
of the expected count matrix $\mathbf{C}_{\rho}^{\tau}$ to the corresponding
singular value $\sigma_{r}(\overline{\mathbf{S}}^{\tau})$ of the
empirical count matrix $\overline{\mathbf{S}}^{\tau}$, as a function
of the lag time. As the small singular values decay quickly with the
lag time, they are dominated by the noise even for small lag times.
Including these noisy singular values would ruin the results. B: Ratio
between mean value and uncertainty (signal-to-noise ratio) from the
bootstrapping for the seven singular values as a function of the lag
time. The thin black dashed line indicates the cut-off we have used
in applications. Only singular values above this line are included
in the estimation, the number of points above this line corresponds
to the OOM model rank, see Fig. \ref{fig:Fig3-one-d-double_well}
H. \label{fig:Fig4_Singular_Values}}
\end{figure}

The above method is theoretically guaranteed to work whenever the
number of MSM states $N$ is at least equal to the number $M$ of
relaxation processes in Eq. (\ref{eq:SpectralDecompKernel}), and
the count matrix $\mathbf{C}_{\rho}^{\tau}$ is of rank $M$. In the
absence of statistical noise, the model rank $M$ can then be determined
by the number of non-zero singular values of $\mathbf{C}_{\rho}^{\tau}$.
For finite data, the numerical rank of $\overline{\mathbf{S}^{\text{}}}^{\tau}$
is not necessarily equal to $M$, as the singular values can be perturbed
by noise. Classical matrix perturbation theory predicts that small
singular values will be particularly affected by noise, see, e.g.,
Ref. \cite{Stewart:1991aa}, and also Fig. \ref{fig:Fig4_Singular_Values}
A. Including noisy and small singular values can severely affect the
accuracy of the method, most likely due to the presence of the matrix
of inverse singular values in Eqs. (\ref{eq:Definition_F1_F2_0}-\ref{eq:Definition_F1_F2_1}).
Also, we expect small singular values to have little impact on the
dominant spectral and stationary properties of the final OOM, but
this will be backed up by further theoretical investigation. 

Consequently, it seems appropriate to cut off small and statistically
unreliable singular values and select a smaller model rank $\hat{M}<M$
in Eqs. (\ref{eq:Definition_F1_F2_0}-\ref{eq:Definition_F1_F2_1}).
In order to determine the uncertainties of the singular values, we
use the bootstrapping procedure, and we discard all singular values
with a signal-to-noise ratio of less than 10. This has proven to be
a useful choice in all applications presented further below. Figure
\ref{fig:Fig4_Singular_Values} B illustrates this procedure for a
simple model system.

\subsection{Software, Algorithmic Details, and Analysis of Computational Effort}

We close the methods section of this paper by pointing out a few more
details of practical importance. First, while it was convenient for
the theoretical analysis to assume that all trajectories sample the
same number of simulation steps $K$, this is not required (see Appendix
\ref{sec:app_B_simulation_length}). Moreover, we also argue in Appendix
\ref{sec:app_B_simulation_length} that all normalizations in Eqs.
(\ref{eq:Definition_si}-\ref{eq:Definition_S2tau}) and (\ref{eq:Empirical_Mean_Si}-\ref{eq:Empirical_Mean_S2tau})
can be dropped in practice. All of the matrices $\overline{\mathbf{s}},\,\overline{\mathbf{S}}^{\tau},\,\overline{\mathbf{S}}_{r}^{2\tau}$
used in the estimation algorithm can be replaced by integer valued
matrices that simply count the number of visits, transitions and two-step
transitions.

Secondly, we have suggested to use the bootstrapping procedure in
order to estimate uncertainties for the singular values of the count
matrix. One way to realize this is to re-draw trajectories with replacement
from the set of all availbale simulations, and to re-estimate the
count matrix from this modified set of simulations. As individual
simulations are statistically independent, this procedure is theoretically
justified and can also be used to estimate uncertainties of further
derived quantities, like timescales and stationary probabilities.
We used the trajectory-based bootstrapping in all examples shown below.
However, if only a small number of rather long simulations is available,
it may be more practical to re-draw individual transitions from the
set of all available transitions in the data set. Let $T$ denote
the total number of data points, which equals $T=KQ$ for uniform
trajectory length, and Eq. (\ref{eq:total_data}) otherwise. If the
transitions were statistically independent, one could simply re-sample
$T$ transition pairs from the set of all $N^{2}$ possible pairs,
where the probability of drawing the pair $(i,j)$ is given by $\overline{\mathbf{S}}^{\tau}(i,j)$.
In fact, transitions are not statistically independent. Therefore,
we suggest to replace the count matrix $\overline{\mathbf{S}}^{\tau}$
by the effective count matrix described in \cite{Noe:2015aa}, but
it should be noted that this procedure relies on several approximations
and must be improved in the future.

Thirdly, we present an overview of the computational cost of each
step in the estimation algorithm in Table \ref{tab:Computational_Cost}
below, assuming that dense matrix algebra is used in every step. It
is expressed in terms of the total number of data points $T$, the
number of MSM states $N$, the OOM model rank $M$, and the number
of bootstrapping samples $n_{b}$.

\begin{table}[h]
\begin{centering}
\begin{tabular}{|c|c|}
\hline 
\textbf{Operation} & \textbf{Cost}\tabularnewline
\hline 
\hline 
Count Matrix Estimation & $\propto T$\tabularnewline
\hline 
Bootstrapping & $\propto n_{b}TN^{3}$\tabularnewline
\hline 
SVD of $\overline{\mathbf{S}}^{\tau}$ & $\propto N^{3}$\tabularnewline
\hline 
Computation of OOM components & $\hat{\boldsymbol{\sigma}}:\,MN+N^{2}$\tabularnewline
\hline 
 & $\hat{\boldsymbol{\Xi}}:\,N\left(N^{2}M+NM^{2}\right)$\tabularnewline
\hline 
 & $\hat{\boldsymbol{\omega}}:\,\propto M^{3}+NM^{2}$\tabularnewline
\hline 
Transition Matrix $\mathbf{T}_{Eq}^{\tau}$ & $N\left(2M^{2}+M\right)$\tabularnewline
\hline 
\end{tabular}
\par\end{centering}
\centering{}\caption{Analysis of computational effort required by the OOM-based estimation
algorithm, if all operations are performed in dense matrix algebra.
\label{tab:Computational_Cost}}
\end{table}
The first step requires an effort which is linear in the data size
and can be performed efficiently. In most cases, we can also assume
the count matrices $\overline{\mathbf{S}}^{\tau},\,\overline{\mathbf{S}}_{r}^{2\tau}$
to be sparse, and the model rank $M$ to be small. In this case, the
cubic term appearing for the calculation of $\hat{\boldsymbol{\Xi}}$
becomes quadratic, while the contributions of the model rank are small.
The only real bottleneck is the singular value decomposition of $\overline{\mathbf{S}}^{\tau}$,
accounting for the factor $N^{3}$ in the second and third step. As
we generally require all singular values of the count matrix, this
step must be performed using dense matrix algebra, which can be time-consuming.
Future research may provide a method that only requires the computation
of the leading singular values, thus allowing for sparse algebra to
be employed.

Lastly, we note that the MSM correction method described in Section
\ref{subsec:Unbiased_MSM_Estimator} is available as part of the pyemma
package \cite{Scherer:0aa}, version 2.3 or later, see \href{http://pyemma.org}{http://pyemma.org}.

\section{Examples}

\label{sec:Examples}

For each of the following examples, we use the trajectory-based bootstrapping
strategy to determine the OOM model rank. Mean values and standard
errors for the singular values are estimated from $n_{b}=10000$ re-samplings,
singular values with a signal-to-noise ratio of at least 10.0 are
accepted. We also generate error estimates for all quantities derived
from the OOM-based Markov model by trajectory bootstrapping, using
1000 re-samplings. In addition, we compute a conventional Markov model
without OOM-based correction as a comparison.

\subsection{One-dimensional Toy Potential}

\label{subsec:One_d_model_system}

\begin{figure}
\begin{centering}
\includegraphics[height=0.6\paperheight]{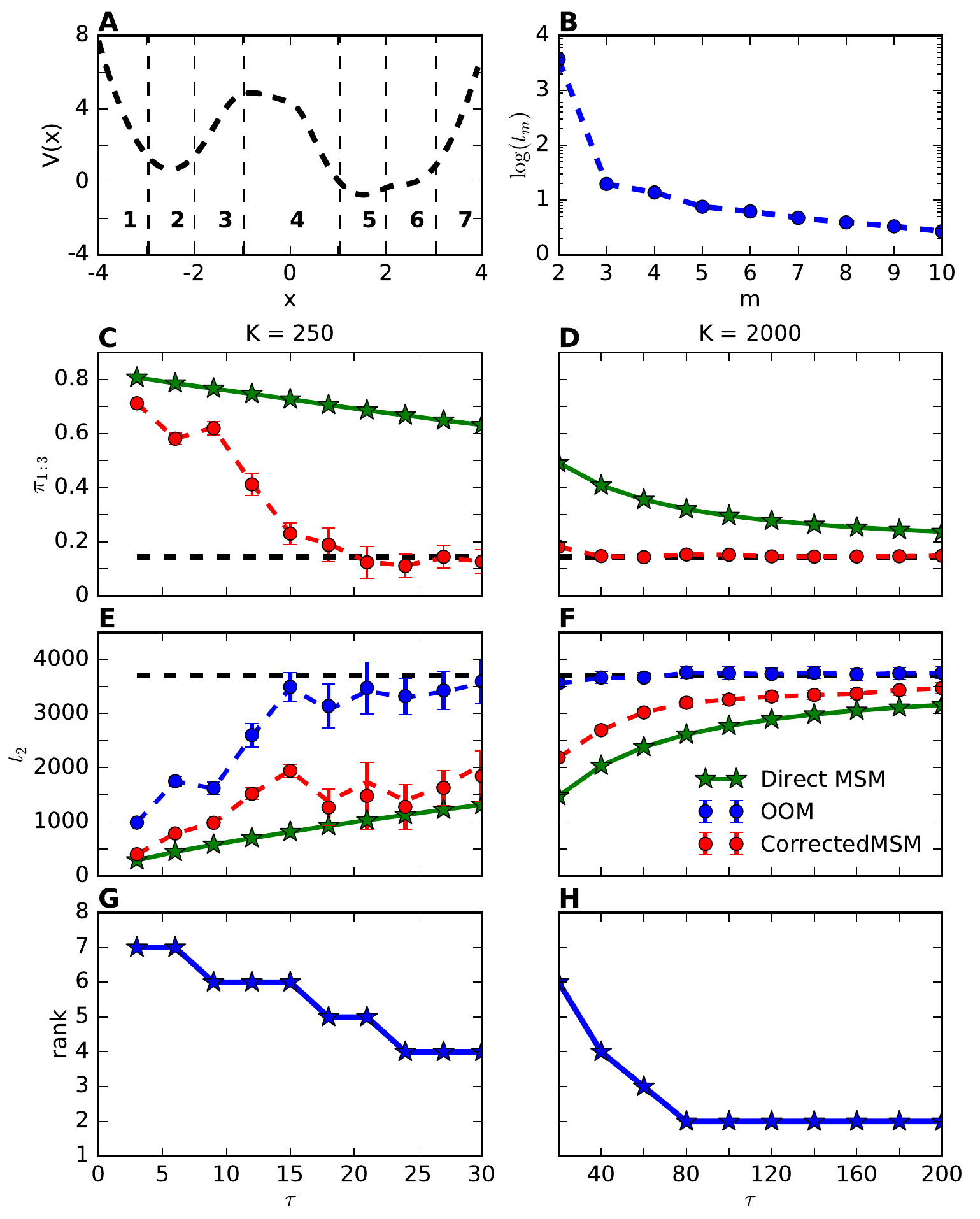}
\par\end{centering}
\caption{A) One-dimensional potential function and discretization of the landscape
into seven states. B) Decadic logarithm of the first nine implied
timescales of the model system. C, D) Estimates of the stationary
probability of states 1-3 from the direct MSM (green) and the corrected
MSM (red), compared to the reference (black dashed line). E, F) Estimates
of the slowest relaxation timescale $t_{2}$ from a direct MSM (green),
the corrected MSM (red) and the OOM-based spectral estimation (blue),
compared to the reference (black dashed line). G, H) Model rank selected
by the bootstrapping procedure. For all quantities derived from the
OOM, the dashed lines indicate the estimated values using the complete
data set, whereas the bullets and errorbars correspond to mean and
standard error from the bootstrapping procedure. Note that errorbars
are hardly visible in panels D and F. \label{fig:Fig3-one-d-double_well}}
\end{figure}

As a first example, we study in more detail the one-dimensional system
used in the introduction. The system is defined by the double-well
potential function shown in Fig. \ref{fig:Fig3-one-d-double_well}
A. The dynamics here is a finite state space Markov chain with $100$
microstates distributed along the $x$-axis, where transitions can
occur between neighboring states based on a Metropolis criterion.
The system is kinetically two-state, as the slowest relaxation timescale
of the system, corresponding to the transition process between the
two wells, is $t_{2}=3708$ steps and clearly dominates all others
(Fig. \ref{fig:Fig3-one-d-double_well}B). 

We investigate the estimation of a seven state Markov model ($N=7$)
using the discretization indicated by dashed lines in Fig. \ref{fig:Fig3-one-d-double_well}
A. Using seven states instead of two accelerates the convergence of
OOM estimates. Still, the seven state discretization is a poor one
- note that state 4 contains large parts of the transition region
as well as parts of the right minimum. This choice was made deliberately
in order to test the robustness of our method with respect to poor
MSM clusterings. We produced two different data sets, each comprising
$Q=5000$ simulations. The first set contains short simulations of
length $K=250$, while the simulations of the second set are $K=2000$
steps long. For the analysis of the smaller data set, we can use lag
times up to $\tau=30$, while we can go to up to $\tau=200$ for the
larger data set. Panels C, E, G of Fig. \ref{fig:Fig3-one-d-double_well}
display the results for the short simulations, while the corresponding
results for the larger data set are shown in panels D, F, H. All simulations
were initiated from a non-equilibrium starting distribution, where
the probabilities to start in each of the seven states are given by
the vector
\begin{eqnarray}
\rho_{1} & = & \begin{bmatrix}0.3 & 0.3 & 0.3 & 0 & 0.05 & 0.05 & 0\end{bmatrix},\label{eq:InitialDist_Oned_System}
\end{eqnarray}
that is, $90$ per cent of the simulations were started in the left
three states, while only $10$ per cent were initialized in the deeper
minimum on the right. Within each state, the actual microstate was
selected from a uniform distribution.

Fig. \ref{fig:Fig3-one-d-double_well}C, D compare estimates of stationary
probabilities from direct MSMs based on Eq. (\ref{eq:Empirical_Mean_TransitionMatrix})
and corrected MSMs with transition matrix given by Eq. (\ref{eq:OOM_Estimator_Ttau_Eq_Rev}).
Due to the non-equilibrium initial distribution, the simulations visit
the left minimum much more frequently than a simulation in equilibrium
would do. While the MSM estimates of the stationary distribution converge
to the true equilibrium distribution at long lag times, they are surprisingly
inaccurate at short times, where the effect of the non-equilibrium
starting distribution still has a strong effect. Even at the largest
lag time $\tau=200$, the bias is still visible. In contrast, the
corrected MSM provides an excellent and stable estimate at lag times
of 15 steps or longer.

In Fig. \ref{fig:Fig3-one-d-double_well}E, F, we compare estimates
of the slowest implied relaxation timescale $t_{2}$ from three different
estimators: A direct Markov model based on Eq. (\ref{eq:Empirical_Mean_TransitionMatrix}),
the corrected Markov model based on Eq. (\ref{eq:OOM_Estimator_Ttau_Eq_Rev}),
and the OOM-based spectral estimation Eqs. (\ref{eq:EV_Decomp_E_Omega_0}-\ref{eq:EV_Decomp_E_Omega_1}).
First, we notice that the direct and corrected MSMs provide different
estimates because of the combination of non-equilibrium starting points
and the poor discretization quality. The corrected MSM timescales
converge faster to the true timescales than the uncorrected ones.
Second, the OOM-based direct estimation of relaxation timescales by
Eq. (\ref{eq:EV_Decomp_E_Omega_1}) provides accurate results already
at lag time $\tau=15$, which is a regime where the number of relevant
relaxation processes cannot be easily determined by a timescale separation,
see again panel B of Fig. \ref{fig:Fig3-one-d-double_well}. The OOM
timescale estimates become very accurate for larger lag times if more
data can be used. Third, the large deviation between the corrected
MSM and the OOM timescales are indicative of the poor discretization
quality employed here. 

Finally, in Fig. \ref{fig:Fig3-one-d-double_well}G, H we show the
model rank selected by the bootstrapping procedure as a function of
the lag time. We can observe how our criterion based on statistical
uncertainties helps to select an appropriate model rank for each lag
time, even when it is not obvious from the timescale plot. As expected,
the system becomes effectively of rank 2 for lag times $\tau\geq100$.

\subsection{Molecular Dynamics Simulations of Alanine Dipeptide}

\label{subsec:ala2_example}

\begin{figure}
\begin{centering}
\includegraphics{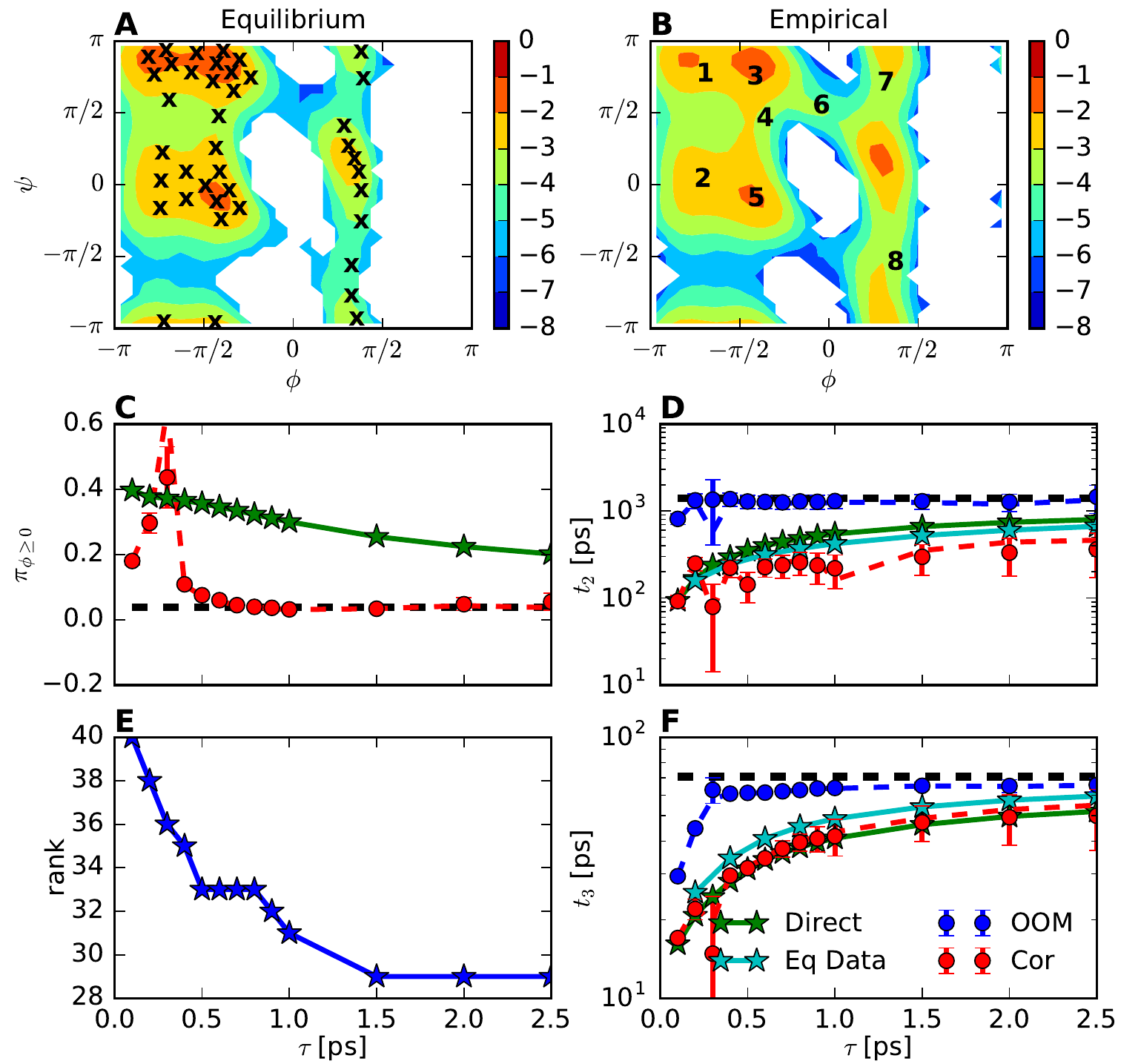}
\par\end{centering}
\caption{Results for alanine dipeptide. A) Equilibrium distribution (logarithmic
scale) in the space of backbone dihedral angles $\phi,\,\psi$ and
clustercenters of a fourty state kmeans discretization used to analyze
the data. B) Empirical distribution (logarithmic scale) sampled by
the data initiated from eight starting structures indicated by the
numbers 1-8. C) Equilibrium probability of all states in the right
part of the plane estimated from the direct MSM (green) and the corrected
MSM (red). Reference in black. D) Estimates for the slowest relaxation
timescale $t_{2}$ from a direct MSM (green), the corrected MSM (red)
and the OOM-based estimation (blue). Reference values from equilibrium
simulations are displayed in black. We also show the expected timescale
estimate using the same fourty state discretization if equilibrium
data was used (cyan line). E) Model rank used for the OOM estimation
as determined by the bootstrapping. F) The same as D) for the second
slowest timescales $t_{3}$. For all quantities derived from the OOM,
the dashed lines indicate the estimated values using the complete
data set, whereas the bullets and errorbars correspond to mean and
standard error from the bootstrapping procedure. Note that errorbars
are hardly visible in panels C and F.\label{fig:fig6-ala2-results}}
\end{figure}

Our second example is molecular dynamics simulation data of alanine
dipeptide (Ac-A-NHMe) in explicit water. Alanine dipeptide has been
used as a model system in numerous previous studies, see Refs. \cite{Nuske:2014aa,Vitalini:2015aa}
and many others. It is well-known that its dynamics can be described
by the two-dimensional space of backbone dihedral angles $\phi,\,\psi$.
Figure \ref{fig:fig6-ala2-results}A shows the equilibrium probability
distribution in this space with its three metastable minima in the
upper left, central left and central right part of the plane. The
slow dynamics consists of exchanges between the left and right part
($t_{2}\approx1400\,\mathrm{ps}$) and between the two minima on the
left ($t_{3}\approx70\,\mathrm{ps}$). We study the estimation of
a Markov model using the discretization also indicated in panel A
of Fig. \ref{fig:fig6-ala2-results}. It was generated by kmeans clustering
of the data set described below using $N=40$ clustercenters. We produced
an ensemble of roughly $11000$ very short simulations of length $20\,\mathrm{ps}$
each. Simulations were initiated from eight different starting structures
labelled by the numbers 1-8 in Fig. \ref{fig:fig6-ala2-results}B,
see Appendix \ref{sec:app_A_ala2} for details. It can be seen that
the resulting empirical distribution does not even reach local equilibrium
within the three metastable regions.

Like in the previous example, we find that it is possible to obtain
precise estimates of stationary probabilities as soon as convergence
of the OOM-based timescales is achieved. In panel C of Fig. \ref{fig:fig6-ala2-results},
we compare results for the equilibrium probability of all states in
the right part of the plane, from a direct MSM and the corrected MSM.
For lag times $\tau\geq500\,\mathrm{fs}$, we are able to correct
the bias introduced by strong non-equilibrium sampling. 

In panels D and F of Fig. \ref{fig:fig6-ala2-results}, we present
estimates of the two slowest timescales $t_{2},\,t_{3}$ produced
by the same estimators as before (OOM in blue, direct MSM in green
and corrected MSM in red). Additionally, the cyan lines correspond
to the timescale estimates of an MSM using equilibrium simulations
and the same discretization (see Appendix \ref{sec:app_A_ala2}).
We find that the OOM-based spectral estimation provides accurate timescale
estimates for short lag times starting at $\tau=500\mathrm{\,fs}$.
Moreover, we notice that for lag times as small as these, MSM timescales
are clearly lower than the true timescales, although a decent discretization
is employed. The difference between OOM and MSM estimates indicates
that an even finer discretization would be required to match the references
at these lag times. The direct estimates, the reference equilibrium
timescales, and our OOM-based estimates of equilibrium timescales,
are nearly identical. Only the mean values extracted from bootstrapping
for $t_{2}$ seem to be a bit low. This will be investigated further.

Finally, the selected model ranks shown in Fig. \ref{fig:fig6-ala2-results}E
confirm that our framework can work in situations where low-rank descriptions
of the dynamics using only a few processes are not adequate.

\subsection{Two-dimensional model system with poor discretization}

\begin{figure}
\begin{centering}
\includegraphics[height=0.6\paperheight]{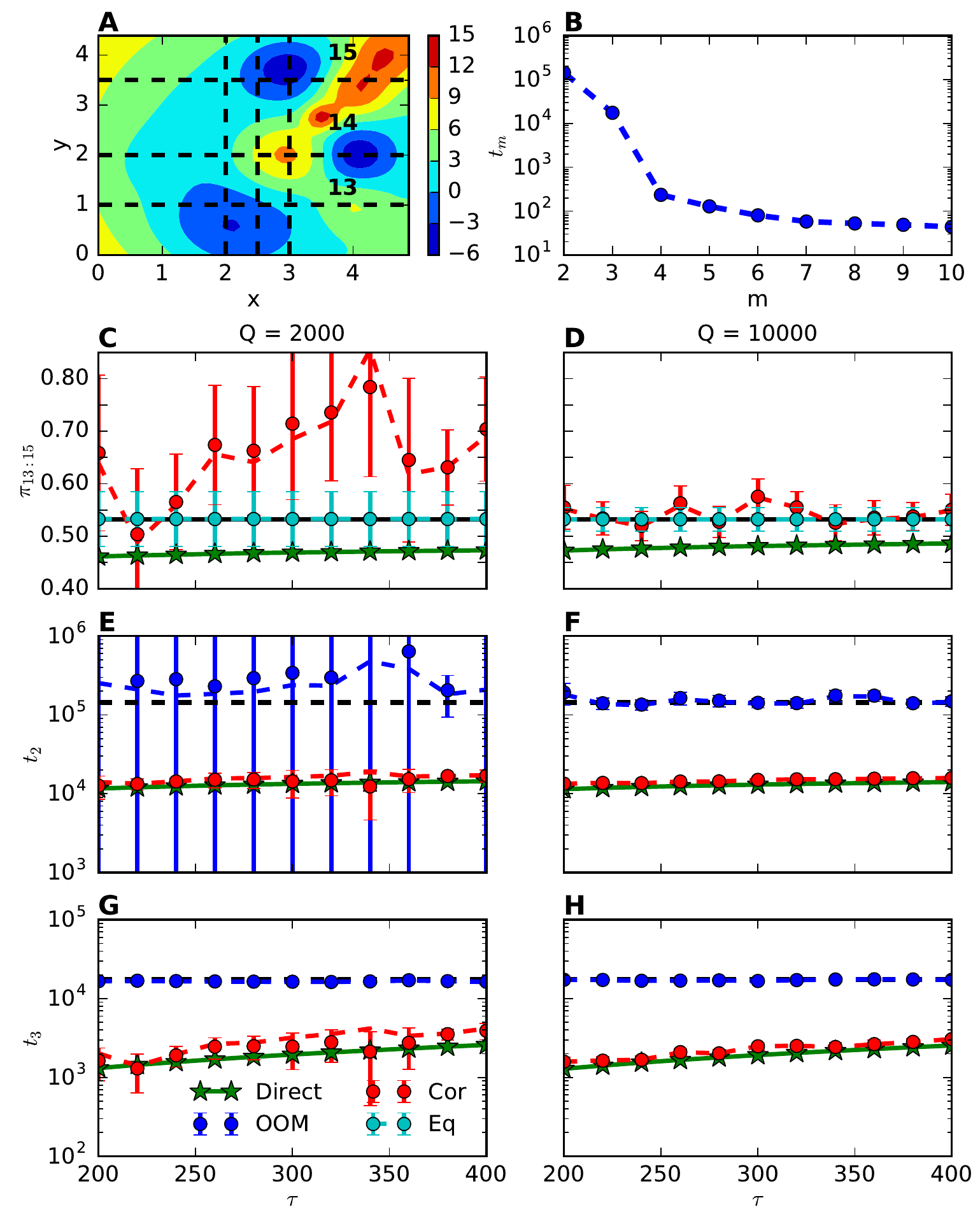}
\par\end{centering}
\caption{A) Two-dimensional potential function with discretization into 16
MSM states indicated by dashed lines. B) Leading nine implied timescales
$t_{m}$ of the system. C, D) Estimates of equilibrium probability
of states 13, 14 and 15 from direct MSM (green) and the corrected
MSM (red), compared to the reference (black line) and estimates from
$100$ different equilibrium simulations, shown by the cyan lines.
E, F) Estimates of slowest relaxation timescale $t_{2}$ from a direct
MSM (green), the corrected MSM (red) and the OOM-based spectral estimation
(blue), compared to the reference (black dashed line). G, H) The same
for $t_{3}$. For all quantities derived from the OOM, the dashed
lines indicate the estimated values using the complete data set, whereas
the bullets and errorbars correspond to mean and standard error from
the bootstrapping procedure. Note that errorbars are hardly visible
in panels F and H. \label{fig:Fig7-two-d-system-1}}
\end{figure}

Our final example is another finite state space Markov chain in the
two-dimensional energy landscape shown in Fig. \ref{fig:Fig7-two-d-system-1}A,
defined by $40\times40$ microstates. Here we show the behavior of
different estimators in an extreme case, where the discretization
is so poor that MSM estimates fail completely. Transitions between
neighboring states are now possible in both $x$- and $y$-direction,
again based on a Metropolis criterion. We study the estimation of
a Markov model using a discretization into 16 MSM states, also shown
in Fig. \ref{fig:Fig7-two-d-system-1}A. As can be seen in Fig. \ref{fig:Fig7-two-d-system-1}B,
there are two dominant timescales, $t_{2}\approx144000$ steps and
$t_{3}\approx17000$ steps. The next timescale is clearly separated
from the first two, after that, there is no more apparent timescale
separation. This time, we fix the simulation length at $K=5000$ steps,
i.e. the trajectories are approximately 30 times shorter than the
slowest timescale. The simulations are started from a uniform distribution
over all microstates. In panels C-H of Fig. \ref{fig:Fig7-two-d-system-1},
we display the results if the number of simulations is set to $Q=2000$
(C, E, G) and $Q=10000$ (D, F, H).

In Fig. \ref{fig:Fig7-two-d-system-1}C, D,, we show the estimation
results for the equilibrium probability of the states labeled 13,
14 and 15. We expect it to be difficult to estimate this probability,
as the states are blending different metastable regions and transition
regions. It can be observed that the estimation of stationary probabilities
is more sensitive to noise, see the results for $Q=2000$. This observation
is not surprising, as the stationary probabilities require accurate
estimation of the two-step count matrices Eq. (\ref{eq:Definition_S2tau})
from the data, which can be more difficult for rarely visited states.
Still, for $Q=10000,$ a reliable estimate is achieved and the biased
estimate of the direct MSM can be corrected. Another comparison we
make is between the estimates from the corrected MSM and those from
long equilibrium simulations that use the same number of total data
points, i.e. $K=2000\cdot5000=10^{7}$ for $Q=2000$ and $K=10000\cdot5000=5\cdot10^{7}$
for $Q=10000$. We show mean values and standard errors from roughly
400 long simulations for $Q=2000$, and roughly 900 simulations for
$Q=10000$. In both cases, the estimates from long equilibrium trajectories
provide more accurate estimates. In practice, however, one needs to
strike a balance between long trajectories that are more beneficial
for the analysis, and short trajectories that can be more efficient
for sampling and state exploration \cite{Preto:2014aa,Doerr:2014aa,Doerr:2016aa}.

Again, we also compare the estimates for the slowest timescales $t_{2}$
(E-F) and $t_{3}$ (G-H) from a direct MSM, the corrected MSM and
the OOM-based spectral estimation. In both cases, correct estimates
of both timescales can be obtained from the OOM, while both the direct
and corrected MSMs estimate timescales one order of magnitude too
small. This suggests that for a bad enough discretization, correcting
for the effect of the non-equilibrium starting distribution will not
be sufficient to achieve convergence in the timescales. However, the
poor discretization quality is revealed by a large error between the
OOM-based estimate and the corrected MSM, and this observation can
be exploited in order to improve the discretization and repeat the
analysis.

\section{Conclusions}

We have investigated the quality of Markov state models when estimated
from many simulations of short length, initiated from non-equilibrium
starting conditions. We have derived an expression for the error between
unbiased MSM transition probabilities and the expected estimate from
many short simulations. This error is shown to depend on the simulation
length, the lag time and the state discretization. If ultra-long trajectories
are employed, i.e. trajectories that are long compared to the slowest
relaxation timescales, then the effect of the initial distribution
is negligible and no further correction is needed. For ensembles of
short trajectories, the situation is more complex. Preparing simulation
trajectories in such a way that they emerge from a local equilibrium
distribution does not appear to be of much practical use: this would
only correct the first transition count of every trajectory while
the subsequent trajectory segments are still biased. The local equilibrium
will be lost for intermediate times along the trajectory as the trajectory
ensemble is not in global equilibrium. In a similar sense discarding
initial simulation fragments can reduce the bias, but cannot systematically
remove it. In particular, since the effect of the bias disappears
with the slowest relaxation times of the system, discarding pieces
of simulation trajectories appears more harmful in terms of reducing
the statistics than it is useful to reduce the bias. With the standard
MSM estimator, the most effective and simplest method to reduce the
bias from the initial trajectory distribution in fact seems to be
using a longer lag time or a better state space discretization. These
are already the usual objectives of MSM construction. However, if
the discretization is poor, the estimation bias due to an non-equilibrium
distribution can be dramatic at practically usable lag times.

The main result of this paper is that we propose an improved estimator
of the MSM transition matrix which is not biased by the initial distribution.
This new estimator is based on theory of observable operator models.
In contrast to the standard MSM estimator, the corrected MSM estimator
does not only use the number of transitions observed between pairs
of states at lag time $\tau$, but also the number of transitions
at lag time $2\tau$. These statistics are combined to get a transition
matrix estimate at lag time $\tau$ that is unbiased by the initial
trajectory distribution. While it may seem that having to estimate
statistics at $2\tau$ is a deficiency compared to standard MSM estimation
when only short simulation trajectories are available, please note
that the corrected MSM estimator can get significantly better estimates
at short lag times, so in practice the lag times needed for a converged
MSM will be smaller than for the standard estimator. 

Finally, we report a result from the OOM framework that shows how
the model-free relaxation timescales can be computed from the same
statistics used for the corrected MSM estimator (i.e. transition matrices
at lag times $\tau$ and $2\tau$). These estimates are only impaired
by statistical error, but are not affected by systematic MSM error
as no MSM is used in the process of obtaining them. The difference
between the corrected MSM timescales and the OOM timescales can be
used in order to assess the discretization quality, as this difference
goes to zero in the limit of good discretization.

This paper addresses the long-standing controversy about the correct
use of simulation data from short non-equilibrium simulations for
MSM estimation, and their effect on the estimation of equilibrium
expectations and kinetics.
\begin{acknowledgments}
This work was funded by Deutsche Forschungsgemeinschaft through SFB
958, SFB 1114 and by the European Commission through ERC starting
grant ``pcCell''. CC is supported by National Science Foundation
(CHE-1265929) and the Welch Foundation (C-1570).
\end{acknowledgments}

\appendix

\section{Simulation Setup of Alanine Dipeptide}

\label{sec:app_A_ala2}

Molecular dynamics simulations of alanine dipeptide in explicit water
at temperature $300\,\mathrm{K}$ were generated with AceMD \cite{Harvey:2009aa}
software using the AMBER ff-99SB-ILDN force field \cite{Lindorff-Larsen:2010aa}
and an integration time step of $2\,\mathrm{fs}$. The peptide was
simulated inside a cubic box of volume $(2.7222\,\mathrm{nm})^{3}$
containing 651 TIP3P water molecules. The Langevin thermostat was
used. The electrostatics were computed every two time steps by the
particle-mesh Ewald (PME) method \cite{Darden1993}, using real-space
cutoff $0.9\,\mathrm{nm}$ and grid spacing $0.1\,\mathrm{nm}$. All
bonds between hydrogens and heavy atoms were constrained.

We have produced $11388$ ultra short simulations of length $20\,\mathrm{ps}$
each, with $50\,\mathrm{fs}$ saving interval. The simulations were
initiated from eight different structures, their projections into
$\phi-\psi$-space are indicated by the number 1-8 in Fig. \ref{fig:fig6-ala2-results}
B. The probabilities to start in each of these structures are given
by the vector

\begin{eqnarray}
\rho_{1} & = & \begin{bmatrix}0.05 & 0.05 & 0.2 & 0.2 & 0.2 & 0.1 & 0.1 & 0.1\end{bmatrix}.\label{eq:Inital_vector_ala2}
\end{eqnarray}
These simulations were used to perform the analyses described in Sec.
\ref{subsec:ala2_example}. Using the same setup, we produced 2363
long runs of $1\,\mathrm{ns}$ simulation time each, with $1\,\mathrm{ps}$
saving interval. We estimated a Markov model on the 40-state kmeans
discretization at lag time $\tau=100\,\mathrm{ps}$ using this data
set, and extracted the reference timescales and equilibrium probabilities
shown as black lines in Fig. \ref{fig:fig6-ala2-results}. Also, we
used the stationary probabilities estimated from this model to initialize
203 short equilibrium runs of $500\,\mathrm{ps}$ simulation time
each, with $100\,\mathrm{fs}$ saving interval. This data set was
used to compute the equilibrium timescales of the kmeans discretization
shown as cyan lines in Fig. \ref{fig:fig6-ala2-results} D, F.

\section{OOM Probability of Observation Sequence}

\label{sec:Appendix_OOM_Path_Prob}

Here, we show the derivation of the path probability formula Eq. (\ref{eq:Path_Prob_OOM}),
that can also be found in Ref. \cite{Wu:2015aa}. In general, the
left-hand side of Eq. (\ref{eq:Path_Prob_OOM}) can be expressed by
repeated integrals over the transition kernel:

\begin{eqnarray}
\mathbb{P}(X_{\tau}\in A_{1},\ldots,X_{l\tau}\in A_{l}) & = & \int_{\Omega}\int_{A_{1}}\ldots\int_{A_{l}}\mathrm{dx_{0}}\ldots\mathrm{dx_{l}}\,\pi(x_{0})p(x_{0},x_{1};\tau)\ldots p(x_{l-1},x_{l};\tau).\label{eq:Path_Prob_Integral}
\end{eqnarray}
Note that $\pi$ appears in the first integral as we assumed that
the dynamics is in equilibrium, i.e. the initial distribution equals
$\pi$. Next, we replace all transitions kernels by the expansion
in Eq. (\ref{eq:SpectralDecompKernel}):

\begin{eqnarray}
\mathbb{P}(X_{\tau}\in A_{1},\ldots,X_{l\tau}\in A_{l}) & = & \sum_{m_{0}=1}^{M}\sum_{m_{1}=1}^{M}\ldots\sum_{m_{l-1}=1}^{M}\left[\int_{\Omega}\mathrm{dx_{0}}\,\pi(x_{0})\psi_{m_{0}}(x_{0})\right]\lambda_{m_{0}}(\tau)\label{eq:Path_Prob_Sum_0}\\
 &  & \left[\int_{A_{1}}\mathrm{dx_{1}}\,\psi_{m_{0}}(x_{1})\pi(x_{1})\psi_{m_{1}}(x_{1})\right]\ldots\lambda_{m_{l-1}}(\tau)\left[\int_{A_{l}}\mathrm{dx_{l}}\,\psi_{m_{l-1}}(x_{l})\pi(x_{l})\right]\nonumber \\
 & = & \sum_{m_{0}=1}^{M}\sum_{m_{1}=1}^{M}\ldots\sum_{m_{l-1}=1}^{M}\delta_{1,m_{0}}\boldsymbol{\Xi}_{A_{1}}(m_{0},m_{1})\ldots\boldsymbol{\Xi}_{A_{l}}(m_{l-1},1).\label{eq:Path_Prob_Sum_1}
\end{eqnarray}
In the second equation, we have used the $\pi$-orthogonality of the
eigenfunctions $\psi_{m_{0}}$ and the fact that $\psi_{1}\equiv1$
in order to replace the $x_{0}$-integral by $\delta_{1,m_{0}}$.
For the last integral, we have also used that $\psi_{1}\equiv1$.
This is a sequence of matrix-vector products. It remains to use $\delta_{1,m_{0}}=\boldsymbol{\omega}(m_{0})$
and that $\boldsymbol{\Xi}_{A_{l}}(m_{l-1},1)=\left[\boldsymbol{\Xi}_{A_{l}}\boldsymbol{\sigma}\right](m_{l-1})$.
In matrix notation, Eq. (\ref{eq:Path_Prob_OOM}) follows:

\begin{eqnarray}
\mathbb{P}(X_{\tau}\in A_{1},\ldots,X_{l\tau}\in A_{l}) & = & \boldsymbol{\omega}^{T}\boldsymbol{\Xi}_{A_{1}}\ldots\boldsymbol{\Xi}_{A_{l}}\boldsymbol{\sigma}.\label{eq:Path_Prob_Sum_2}
\end{eqnarray}
Finally, note that this derivation also works if the dynamics is not
in equilibrium. In this case, the vector $\boldsymbol{\omega}$ is
given by $\boldsymbol{\omega}(m_{0})=\int_{\Omega}\mathrm{dx_{\text{0}}}\,\rho_{0}(x_{0})\psi_{m_{0}}(x_{0})$,
where $\rho_{0}$ is the non-equilibrium initial condition.

\section{Variable Simulation Length}

\label{sec:app_B_simulation_length}

Here, we verify that the estimation algorithm from Sec. \ref{subsec:Unbiased_MSM_Estimator}
can be applied to data sets comprised of simulations of non-uniform
length. We assume that for $j=1,\ldots,J$, there is an ensemble of
$Q_{j}$ simulations of length $K_{j}+2\tau$, i.e. $K_{j}$ transition
pairs / triples will be used from each of these trajectories. We assume
that $Q_{j}\rightarrow\infty$ for all $j$, s.t. every sub-ensemble
samples from an empirical distribution $\rho_{j}$. Define the number
of data points generated by the $j$-th ensemble as $T_{j}=Q_{j}K_{j}$,
and the total number of data points by

\begin{eqnarray}
T & := & \sum_{j=1}^{J}Q_{j}K_{j}.\label{eq:total_data}
\end{eqnarray}

Moreover, we assume that $\frac{T_{j}}{T}\rightarrow\alpha_{j}$,
i.e. the fraction of data points generated by the $j$-th ensemble
approaches a constant for all $j$. Let us define the distribution

\begin{eqnarray}
\rho & = & \sum_{j=1}^{J}\alpha_{j}\rho_{j}.\label{eq:Full_Ensemble_Variable_Length}
\end{eqnarray}

Trajectories of length $K_{j}+2\tau$ are enumerated by $q_{j}$ and
labelled $\mathbf{Y}_{q_{j}}$. Further, let $s_{K_{j}}(\mathbf{Y}_{q_{j}})$
be any of the estimators from Eqs. (\ref{eq:Definition_si}-\ref{eq:Definition_S2tau}),
where the subscript $K_{j}$ indicates that $K-2\tau$ in Eqs. (\ref{eq:Definition_si}-\ref{eq:Definition_S2tau})
must be replaced by $K_{j}$. In addition, denote by $s(\mathbf{Y}_{q_{j}})$
the same estimator, but without the normalization. Also, let $c_{\rho_{j}}$
denote the corresponding correlation from Eqs. (\ref{eq:Definition_Ctau}-\ref{eq:Definition_ci})
and (\ref{eq:Definition_C2tau}) w.r.t. the density $\rho_{j}$. It
follows that
\begin{eqnarray}
\overline{s}_{T} & := & \frac{1}{T}\left[\sum_{q_{1}=1}^{Q_{1}}s(\mathbf{Y}_{q_{1}})+\ldots+\sum_{q_{J}=1}^{Q_{J}}s(\mathbf{Y}_{q_{J}})\right]\label{eq:Variable_Length_Estimator_0}\\
 & = & \frac{T_{1}}{T}\left[\frac{1}{T_{1}}\sum_{q_{1}=1}^{Q_{1}}s(\mathbf{Y}_{q_{1}})\right]+\ldots+\frac{T_{J}}{T}\left[\frac{1}{T_{J}}\sum_{q_{J}=1}^{Q_{J}}s(\mathbf{Y}_{q_{J}})\right]\\
 & = & \frac{T_{1}}{T}\left[\frac{1}{Q_{1}}\sum_{q_{1}=1}^{Q_{1}}s_{K_{1}}(\mathbf{Y}_{q_{1}})\right]+\ldots+\frac{T_{J}}{T}\left[\frac{1}{Q_{J}}\sum_{q_{J}=1}^{Q_{J}}s_{K_{J}}(\mathbf{Y}_{q_{J}})\right]\label{eq:Variable_Length_Estimator_1}\\
 & \rightarrow & \alpha_{1}\mathbb{E}\left(s_{K_{1}}\right)+\ldots+\alpha_{J}\mathbb{E}\left(s_{K_{J}}\right)\label{eq:Variable_Length_Estimator_2}\\
 & = & \alpha_{1}c_{\rho_{1}}+\ldots+\alpha_{J}c_{\rho_{J}}\label{eq:Variable_Length_Estimator_3}\\
 & = & c_{\rho}.\label{eq:Variable_Length_Estimator_4}
\end{eqnarray}
The convergence in Eq. (\ref{eq:Variable_Length_Estimator_2}) is
convergence in probability. Thus, if we sum up all visits / transitions
/ two-step transitions, and divide by the total number of data points
in the end, we arrive at an asymptotically correct estimator of the
correlations w.r.t. the density $\rho$. As the OOM estimation algorithm
only relies on consistent estimators for correlations w.r.t. some
empirical density $\rho$, it can still be applied in this setting.
Finally, the normalization by $\frac{1}{T}$ can be omitted in practice,
because it cancels out in Eqs. (\ref{eq:Estimator_E(S_i)}-\ref{eq:OOM_Estimator_pi_0}).

\bibliographystyle{unsrt}

\end{document}